\documentclass[final,3p,times]{elsarticle}

 \usepackage{graphicx}
 \DeclareGraphicsExtensions{.pdf,.png,.jpg}

\usepackage{color}
\usepackage{float}
\usepackage{verbatim}

\usepackage{amsmath}
\usepackage{PNotation_art}

\usepackage{listings}

\usepackage{amssymb}
\usepackage{amsthm}

\date{August 9, 2016}
\journal{Future Generation of Computer Systems}

\begin{document}

\begin{frontmatter}



\title{Model-Driven Development of Data Intensive Applications over Cloud Resources}
\tnotetext[t1]{This work was co-financed by the Industry and Innovation department of the Aragonese Government and European Social Funds (COSMOS research group, ref. T93); and by the Spanish Ministry of Economy under the program ``Programa de I+D+i Estatal de Investigaci\'on, Desarrollo e innovaci\'on Orientada a los Retos de la Sociedad'', project identifier TIN2013-40809-R }

\author[unizar]{Rafael Tolosana-Calasanz\corref{cor1}}
\author[unizar]{Jos\'e \'Angel Ba\~nares}
\author[unizar]{Jos\'e-Manuel Colom}
\cortext[cor1]{Corresponding author}

\address[unizar]{Computer Science and Systems Engineering Department\\Arag\'on
Institute of Engineering Research (I3A)\\University of Zaragoza, Spain}

\begin{abstract}
The proliferation of sensors over the last years has generated large amounts of raw data, forming data streams that need to be processed.

In many cases, cloud resources are used for such processing, exploiting their
flexibility, but these sensor streaming applications often need to support operational and control actions that have real-time and low-latency requirements that go beyond the cost effective and flexible solutions supported by existing cloud frameworks, such as Apache Kafka, Apache Spark Streaming, or Map-Reduce Streams. 
%Furthermore, there are many ways in which a streaming application can be mapped onto cloud resources, providing the same functionality but with different concurrent behaviours.
In this paper, we describe a model-driven and stepwise refinement methodological approach for streaming applications executed over clouds. The central role is assigned to a set of Petri Net models for specifying functional and non-functional requirements. They support model reuse, and a way to combine formal analysis, simulation, and approximate computation of minimal and maximal boundaries of non-functional requirements when the problem is either mathematically or computationally intractable. We show how our proposal can assist developers in their design and implementation decisions from a performance perspective. Our methodology allows to conduct performance analysis: The methodology is intended for all the engineering process stages, and we can (i) analyse how it can be mapped onto cloud resources, and (ii) obtain key performance indicators, including throughput or economic cost, so that developers are assisted in their development tasks and in their decision taking. In order to illustrate our approach, we make use of the pipelined wavefront array.

\end{abstract}


\end{frontmatter}

\section{Introduction}
\label{sec:intro}
%Kind of applications
The confluence of cloud technologies, Internet of Things (IoT) sensors, and big data analytics has been recognised as the key enabling combination of technologies for innovation in a broad range of sectors, including military applications, the emerging smart grids, smart buildings, applications for e-health, or natural disaster prevention.

%Characterization of app CDFA
A common characteristic of all these applications is that they are data intensive, with data {\color{black} being generated continuously and coming from heterogeneous sources such as sensors or scientific devices. Furthermore, data generation rates can vary significantly, and the applications may often need to process data in a timely manner enabling systems to take corrective / strategic operational actions, or react at situations~\cite{Mazmanov:2013aa} urgently. For this reason, such applications make often use computational resources. We will refer to this kind of applications as continuous, data flow applications (CDFA), involving a wide range of applications, such as scientific workflows, pipelines, streaming applications, or any other data intensive application where data dependencies and concurrency play an important aspect. 

It is difficult to have one solution valid for all these applications in any circumstance. An approach for their conception can make use of any of the existing, enabling framework. Some representative examples can be found in commercial clouds (e.g. Amazon, Google, or Microsoft), or in any of the more than 40 projects of the Apache Big Data Stack~\cite{Fox:2015zh}, which include the pioneer MapReduce computation framework, or others such as Flume, Spark, Storm, or Flink.
 
 %Enabling technologies. Problem to focus soon on technologies
Such solutions often provide high-level operators that hide the inherent complexity of a distributed resources: Programmers can make use of the operators, but they typically do not contemplate the computational resource usage in their functional implementation (see the Spark implementation in Section~\ref{sec:req}).The availability of this rich enabling ecosystem can make the design process to abandon too soon the analysis to produce an implementation, avoiding a deep investigation of properties, or a wider exploration of alternative architectural designs and task mapping solutions onto distributed resources. This lack of analysis may result in some areas of the design space being insufficiently explored. In other words, if the analysis is done without an enabling technology in mind, a variety of alternative implementations might emerge~\cite{Mattson:2004cs}.

Indeed, lifting the level of abstraction is the most effective way to manage problem complexity. In order to illustrate this statement, it is easy to see that programmers do not need to worry about writing parallel code~\cite{Abadi:2003nb,Gulisano:2010fl, Lohrmann:2014jt}, or about the cluster of machines where the application will be finally executed, i.e. these aspects can be abstracted away from the high-level design process of the application. However, since there is a strong relationship between the application logic and the execution environment that can affect its functionality and disturb the non-functional properties to be fulfilled, it is advisable to contemplate the execution environment in the abstraction of the system.}

{\color{black}
This paper is about how to manage the complexity of developing the logic of data intensive applications; initially, by taking into account the functional and non-functional requirements, to gradually incorporate the restrictions imposed by the implementation in later phases --a typical non-functional requirement can be scalability. An application can scale if it can exploit concurrency and data dependencies, which require a careful analysis of the application (e.g. via model analysis) to determine the existing opportunities to execute multiple activities at the same time, considering eventually features such as the variability of data and the executing environment.

Hence, our main contributions are linked with the shortcomings, gaps and difficulties found in the development of cost-effective and performance sensitive CDFAs, managing the complexity of developing these applications from three different fronts~\cite{DBLP:journals/computing/AndrikopoulosBLS13}:} (i) Lifting the level of abstraction for applications based on cloud frameworks, (ii) Providing a collection of patterns for parallel applications and cloud platforms that describe high-quality solutions to recurring  problems, and (iii) Developing ad-hoc performance models to forecast the behaviour of particular patterns on specific platforms.

 L. Lamport emphasizes two key ideas related to specifications in~\cite{Lamport:2015aa}: (i) It is better to handle the complexity of applications by abstractions, instead of hiding it; and (ii) specifications should not be written in program code. For an engineer designing a CDFA, these ideas can be translated to consider resource aware functional specifications  instead of pure functional specifications; and avoid functional specifications that are written in the abstractions provided by a concrete implementation framework.

 The incorporation of resources \emph{early} at the functional specification will allow developers to perform an analysis from the beginning that will lead to a wider exploration of the design space. This analysis must help find errors, and to think about concurrency and data dependencies as the main features on which the scalability, performance or economical cost requirements rest. The level of abstraction that requires the incorporation of resources to the functional level is not the same as the detail that might require the implementation on a \emph{particular infrastructure}. It is easier to find and fix errors in higher-level abstractions than in the code of the implementation.  It implies that a hierarchical model is required that can be refined for  different execution platforms.  Finally, the need for reasoning about performance and finding design errors lead to start writing formal specifications. The complexity that arises from the combination of distributed functionality,  performance, and variability of the cloud environment  involves the use of predictive models with different accuracy degrees~\cite{Harrison:1992vn,Khazaei:2012rg}. Summarizing, our research objective is a step-by-step  methodology to guide  developers  for developing specification that incorporates the  abstractions required by a CDFA,  and help them to support the analysis  and design with the use of formal models.

 The general contribution in this paper is a model-driven and stepwise refinement methodological approach for data intensive applications executed in distributed systems.The central role in this methodology is assigned to a set of Petri Net models describing the functionality required and its behaviour from a non-functional perspective. Such Petri Net models allow a developer to analyse the behaviour of the system prior to and during its implementation and deployment. In particular, the analysis enables the exploration of minimal and maximal boundaries of the economic cost of the execution of the application, in relationship with its performance and its workload. Hence, in many cases, as an outcome of the analysis of the system, changes may be recommended or induced into the system with the purpose of modifying parts of the design and enforcing the agreed specifications. Such changes can happen at different levels, i.e. at task specification level or at resource management level. Therefore,  as a by product of the approach, the proposed methodology can improve the design and implementation's decisions taken during the development process of a streaming application.

{\color{black} On the other hand, cloud infrastructures are particularly well-suited for the computations of such applications due to their properties and they have typically represented the technology of choice for the aforementioned current framework projects. In particular, our approach considers the following properties of the cloud: (i) {\em on-demand resource provisioning}, (ii) the {\em dynamism of the cloud} execution environment caused by performance variation of machines, services competing for shared resources, and changing user quality of services requirements~\cite{Yeo:2011cy,OLoughlin:2014ww}. The on-demand resource provisioning characteristic enables us to configure and manage the computational cloud resources so that one of the task-to-resource mapping solutions analysed can be deployed. Furthermore, as a result of our analysis, as already stated, we can explore the alternative mapping solutions for the application tasks and their performance. Then, we can choose the most suitable one and exploit the cloud in order to configure and provision the computational resources at runtime. In case of performance degradation --  it is known that cloud resources are subject to resource contention, leading to performance interference~\cite{nathuji2010q,medel2016modelling} over time, and potentially impacting performance metrics, making the execution time to vary up to an order of magnitude -- our approach enables the application to exploit the versatility and elasticity of cloud infrastructures to migrate into an alternative mapping solution. In this paper, however, our focus is on how to obtain the models and how to perform the subsequent analysis from them so that they can assist developers across the CDFA lifecycle. We are not focused
on how to exploit a cloud infrastructure, but we will provide a discussion about it.

\section{Data-Intensive Applications and Current Technological Practice}
\label{sec:req}
\textcolor{black}{Over the last years, the rapid development of science and engineering is generating a wide variety and large amount of datasets. Such data is growing larger and its location, availability and other properties are often dynamic, that is, dependent on time~\cite{DBLP:journals/concurrency/JhaKLHRS17}.
From a computational perspective, the amount of computational resources required for processing it is typically significant, such datasets often need to be processed within a time threshold; and, as a result of the output generated, automated control actions can be triggered. In this section, we will discuss a number of such applications and their requirements. In particular, we will describe the smart building scenario and we will show an implementation of it in the \emph{Spark framework}. Spark is currently one of the popular frameworks in use, consequently, we believe that its exemplification can be considered as a good representative of current technological practice for this kind of applications.}

\subsection{Motivating Case Studies}
\textcolor{black}{
An example of a data-intensive application can be found in the biological sciences. The study of marine ecosystems requires constant monitoring and analysis of undersea life and for such a purpose undersea video data is available. In many cases, human labor was used for undersea video analysis, but this is a tedious task. The EU-funded Fish4Knowledge project~\cite{www-fish4knowledge} developed algorithms and a distributed infrastructure in order to support automated video analysis. The idea is that videos are recorded and transmitted continuously for processing. Nevertheless, the processing of each video is computationally intensive and also needs to be conducted in a timely manner. On the other hand, in the automotive industry, automobile safety is becoming popular in order to reduce the consequences of traffic collisions. There can be complex scenarios within city boundaries that may require immediate reaction: A vehicle at a relatively low speed of 60 km/h can cover more than 3 meters in 200ms. Vehicles
will incorporate sensors collecting measurements periodically, and the processing needs to respond fast, acceptable delays for collision avoidance systems should be below 10ms~\cite{7403840}.}

\textcolor{black}{Besides, the transformation of power networks into smart grids requires a controlled charging of batteries for electric vehicles~\cite{DBLP:journals/concurrency/Tolosana-Calasanz17}. 
From a computational perspective, such processing requires periodic computations of charging schedules which considers electricity price,
electrical constraints, and user preferences. The computations need to be done within a time threshold, where a breadth-first search algorithm is executed for each geographic area to prioritize on which vehicle should be selected for charging, given that demand exceeds supply).}

\textcolor{black}{Hence, we can conclude that there are a number of applications arising with the proliferation of distributed sensors. Many of them share common characteristics: They are typically computationally intensive, some require immediate response (like in the automobile safety scenario), while some others do not, as data elements arriving into the system need to be processed within a time threshold (deadline). Such a deadline is in the order of seconds, minutes, hours, or even days, rather than in the order of milliseconds, and this deadline is one of the key metrics in the Service Level Agreement (SLA). Moreover, the processing may involve the execution of complex simulations or control algorithms~\cite{DBLP:journals/concurrency/Tolosana-Calasanz17}.}

\subsubsection{The Smart Building Scenario}
\textcolor{black}{
Another similar data-intensive application is the smart building scenario. An advanced intelligent building management aims at reducing operational costs, while increasing the energy saving in large buildings via automated management actions. Therefore, on one hand, some physical variables are measured periodically by sensors deployed across a building infrastructure; such monitored physical variables often include temperature or humidity; and also people density across the building premises. On the other hand, a number of factors of the building are also considered, such as the construction materials, the structure of the building, or the building heat and mass balance. With all such monitored values and characteristics a number of computations can be accomplished, so that an automatic management action is subsequently taken, i.e. for each room, increasing / decreasing its temperature, customising its lighting, etc. In particular, the computations are often based on the EnergyPlus model~\cite{DBLP:conf/ccgrid/PetriRRLBZMP14}, which is a simulation framework. Often the output of such simulations needs to be obtained before a given deadline, in time for taking any required control actions.} 

\subsubsection{Spark and its Programming Model}
\textcolor{black}{Spark~\cite{zaharia2010spark} is a popular cluster computing framework that arose to overcome some limitations of the MapReduce paradigm, since there are some applications whose control / data flow cannot be expressed efficiently, in MapReduce, as acyclic flows: (i) iterative jobs, when a function needs to be applied repeatedly over the same dataset, and (ii) interactive analytics, when exploratory queries need to be applied on large datasets. Spark offers programmers an interface that builds in implicit data parallelism. The execution is specially designed to be fault tolerant. Such interface consists of parallel operations such as map, filter, reduce, or foreach that are accomplised by passing closures (functions) to Spark. All these operations are based on the \emph{resilient distributed dataset} (RDD) abstraction, which represents a read-only collection of objects partitioned across a set of machines that can be rebuilt if a partition is lost.} 

\textcolor{black}{In order to illustrate the usage of Spark, we make use of the aforementioned Smart Building scenario and we code the EnergyPlus based simulation scenario based on the computations specified at~\cite{DBLP:conf/ccgrid/PetriRRLBZMP14}. The following code fragment in Spark \& Java for the Smart Building scenario can be seen in Listing~\ref{cod:spark}. First, a one-second batch Spark context is created, thereby all the requests arriving during the previous second will be processed. Each request has all the required data for a building simulation, namely the monitored data and the building characteristics. Then, after an input stream is created, the operations over the datasets
are specified: (i) for each request (task), a number of simulations are derived; this is achieved by means of the parallel operator \emph{flatmap}, which applies the lambda function generateJobs to each request (task); (ii) finally, each simulation is computed in parallel by means of the \emph{foreachRDD} method. It should be noted that in Spark, developers specify first the data flow, but the actual computations do not start until the method \emph{start} is invoked (line 16 of Listing~\ref{cod:spark}).}

\lstset{numbersep=5pt,numberstyle=\tiny,numbers=left,breaklines=true,language=Java}          % Set your language (you can change the language for each code-block optionally)

\begin{lstlisting}[caption={Smart Building Scenario code in Java / Spark}, label={cod:spark}]  % Start your code-block

  public static void main(String[] args) throws Exception {  
    // 1 Create the context with a 1 second batch size
    SparkConf sparkConf = 
    	new SparkConf().setAppName("SmartBuildingEnergyPlus");
    JavaStreamingContext ssc = 
    	new JavaStreamingContext(sparkConf, Durations.seconds(1));
    // 2 Create an input stream on target ip:port 
    JavaReceiverInputDStream<String> task = 
    	ssc.socketTextStream(args[0], Integer.parseInt(args[1]), 
            StorageLevels.MEMORY_AND_DISK_SER);
    // 3 For each task, create the EnergyPlus jobs
    JavaDStream<Job> jobs = task.flatMap(x ->  EnergyPlus.generateJobs(x));
    // 4 Execute each job: Perform the EnergyPlus simulation
    jobs.foreachRDD();
    // Start the application
    ssc.start();
}
\end{lstlisting}

\textcolor{black}{This example illustrates how the programmer specifies data flow operations in Spark, regardless of the computational resources required. The advantage of the approach is that the inherent complexity of the distributed and parallel systems is hidden from developers. Once the engine starts the execution of parallel operators on the receiving datasets, the computational resources required will be provisioned. Spark can be configured in a number of different ways, for instance, with the standalone Spark resource provisioning component or with Mesos~\cite{hindman2011mesos}. Nevertheless, the disadvantage of the approach is that it can result in an inefficient resource provisioning, which is the essence of this work. Indeed, Spark resource provisioning components cannot operate with the information regarding the characteristics of the data flows (e.g. the workload prediction characteristics, such as arrival rate, predicted execution time, etc.) and consequently any elasticity operation at runtime can be challenging.}

\textcolor{black}{For each building, the Spark scheduler needs to execute the simulations (line 14). This step can be computational intensive and there is a trade-off between having the output in time (i.e. measured in terms of a significant percentage of simulations finished within a given threshold~\cite{DBLP:conf/ccgrid/PetriRRLBZMP14}) and the number of computational resources involved. The amount of computational resources is a key aspect, since too few resources may lead to an SLA violation, but a too high number of resources may incur in an increase in economic cost. However, even in this simple scenario, where jobs (simulations) have no dependencies among them, performance uncertainty may arise, leading to SLA serious violations. Indeed, in line 14 of Listing~\ref{cod:spark}, the scheduler makes a decision on how many computational resource to allocate for the workload, but such decision can seriously affect the future workload to come, as it is completely unknown by the system the amount of
computational resources required for the immediate future, and the choice on the number of used resources need to be done immediately.
The solution adopted in~\cite{DBLP:conf/ccgrid/PetriRRLBZMP14} consists in killing running jobs (simulations) when more resources are needed
and there was no free resource left, thus reducing the EnergyPlus simulation accuracy.}

\textcolor{black}{With our approach, by taking into account performance models
that also consider computational resources, we can better estimate the number of resources required at any step and move from a purely reactive provisioning approach (that can potentially kill jobs) into a predictive one (that can potentially make a more efficient and intelligent use of resources).}

\subsection{Computational Capacity Requirements}
Given all the previous characteristics of the \textcolor{black}{data intensive applications}, and also based on related previous work~\cite{DBLP:journals/concurrency/JhaKLHRS17,DBLP:journals/tpds/Tolosana-Calasanz17}, we have identified the following key computational resource requirements:
\begin{itemize}
\item Support for data-intensive workloads: The type of processing required is computationally intensive and it may involve a great number of computational resources. The computations may require complex distributed and parallel simulations, optimisation and control algorithms, or the computation of predictive models.

\item Quality of Service (QoS) enforcement: Typically, each data element may need to be processed within a time threshold. Sometimes, such
threshold is similar to the processing and transmission times, therefore no delay is allowed (like in the vehicle safety applications). In other cases, the processing time ($S$) is less than the overall due time for the control period (deadline), allowing data elements to be buffered prior to their processing. Moreover, in some scenarios, exceeding the overall amount of time for performing the computation may be allowed by the SLA of the application. The resource management policy should provide mechanisms for enforcing QoS.

\item Elastic/On-demand Provisioning: The computational capacity must be adjusted to the overall requests. Therefore, a resource management strategy needs to allocate the appropriated number of computational resources to process unpredictable and variable workloads while satisfying QoS constraints, as described above. The underlying system must be able to support admission control and enable a variable processing rate per stream. The adopted mechanisms should be based on autonomic principles, so that they are resilient and self-adaptive to variations in the historical traces without requiring human intervention.
\end{itemize}

The cloud computing paradigm and related technologies are appropriate for these computational capacity requirements. Computational resources (i.e.
CPU, network and storage) can be customised for the specific needs of the data intensive workload. Then, vertical / horizontal (de-)provisioning
of resources can be exploited in order to accommodate the computational power to the workload demand, in such a way that QoS is enforced
and the economic cost of the resource usage is minimised. In the following sections, we describe our methodology for building Petri-net based
performance models that can be subsequently exploited to choose the appropriate target cloud paradigm. Our model supports two types of analysis, namely, performance and economic based.

%=================================================================================================

\section{Methodology for the Functional, Performance and Economic Analysis of CDFAs on the Cloud}
\label{sec:mod}

In this section, we outline the  principles for the modular construction of specifications that will guide the analysis, design and execution of CDFAs on the cloud. Application development typically starts with the requirements capture phase of functional, performance, and economic aspects of the problem domain. Any software development methodology must be founded on scientific-based, predictable, and rigorous models to be considered an engineering method.   Formal models support the verification of design, coding and testing phases for finding a mismatch between a system's requirements and its actual implementation~\cite{Holzmann:2007ly,Seceleanu:2013ol,Lamport:2015aa}.
Unfortunately, most computer scientists are either not familiar with or reluctant to use formal methods~\cite{Lamport:2015aa}. Moreover, formal-model based analysis tools are only useful under certain assumptions (e.g. imposing some boundaries). Hence, it has so far been reasonably effective to tune an optimisation. On the other hand, simulation may be useful to discover some (un)desirable behaviours, but in general it does not allow to prove the (in)existence of some properties. Therefore, the combination of simulation and formal models for functional, performance, and economic analysis is necessary for efficient and reliable design and/or optimisation.

\subsection{Requirements for our Methodological Approach}
In order to support the methodology, we have identified the following modelling requirements that go beyond pure functionality: 
\begin{itemize}
\item The methodology should define a {\bf development process}, which is guided by the identified abstraction levels, and should provide a number of modelling artefacts, analytical methods, and guidelines to support it. The guidelines should include a {\em cyclical process} including analysis, design, and testing. It is not necessary to define the whole set of requirements from early stages. The process can be based on a top-down approach, starting with  functional requirements, and following a refinement process with non-functional requirements.  Alternatively, we can start modelling the resource management of a concrete framework to accomplish some non-functional requirements, and following a bottom-up approach specify the functional specification on top.  A mixed {\em bottom-up} and {\em top-down} design is required depending on the level being modelled, which enables the generation of reconfigurable and scalable specifications. 
%R1 Rephrase the term different views.
\item The development of CDFAs must be supported by a {\bf specification  language}  to describe a them as a collection of platform-independent building blocks. There are several streaming processing engines and frameworks based on particular parallel patterns to build and deploy tasks as distributed applications for commodity clusters and clouds.  The language should support complementary capacities for the description of the application: Behavioural specification of concurrent processes,   transformations operated over the data flow, and structural description of components that constitute the application.  The reasonable approach is a general specification language that combines  a {\em high-level language,  describing the concurrency pattern or abstraction} to create very large programs. With such a general specification language, a user can plug in simple sequential programs expressed in whatever language  to create very large parallel programs \cite{Etzion:2010in,C.Pautasso:2006ss,Yu:2010tl, Simmhan:2014ia, Lohrmann:2015rt}.   The specification should also support a formal {\em component-based development}  to build models from existing modules and a capability to reason about the resulting composition.  Hence, the specification must be executable to support both {\em analytical analysis} and {\em simulation}. 
\end{itemize}

\subsection{Synoptical View of the Methodology}
There are many ways in which a system can be built to provide the same functionality  with different concurrent behaviours and different deployments over distributed computing resources.  Starting with the modelling artefacts, our methodological  approach can be summarised with the equation '{\em Specification of Continuous Data Flow Applications (CDFA)} = Functional Entities + Communication / Synchronisation mechanisms + Data Dependencies + Resources'. The identification and characterisation of each building block of the proposed equation defines the basic specification elements. Figure~\ref{Figure0} depicts a synoptical view of our methodology. \textcolor{black}{In contrast to the Spark model of computation discussed in Section~\ref{sec:req}, it} identifies three levels that are interleaved, namely functional, operational, and implementation. 
 %They all involve qualitative and quantitative analysis of models
The overall life-cycle is finalized with implementation and monitoring phases:
\begin{figure}

\includegraphics[width=1\textwidth]{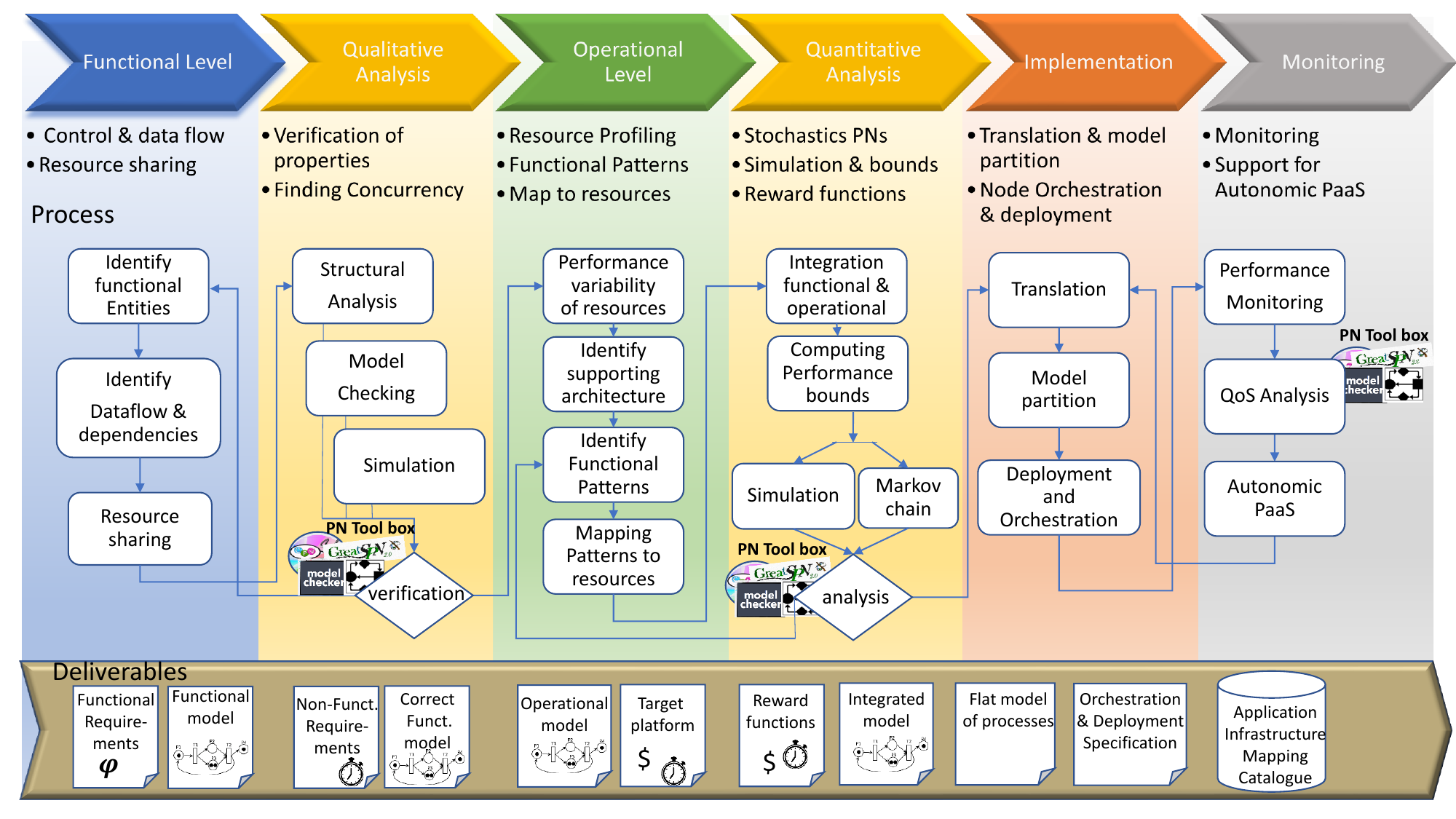}
\centering
\caption{Life cycle and management methodology for CDFAs over cloud platforms. }
\label{Figure0}
\end{figure}

\begin{itemize}
\item {\bf Functional Level} The process starts by identifying the functional requirements of the problem domain and the outcome of this step is a functional model.
\item{\bf Qualitative Analysis} A functional model analysis can identify problems and help guide the redesign of the functional model aiming to achieve the maximum level of concurrency.  
\item  {\bf Operational Level} The operational level takes into account the execution platforms, and it explores the design space to select the design pattern that most effectively defines how to map processes to resources.  The outcome is an operational model.
\item {\bf Quantitative Analysis}  The integration of the functional and the operational model allows the designer to evaluate performance and reward functions.  The analysis can help  guide  the redesign of the functional  and operational models to meet  non-functional requirements.  
\item {\bf Implementation Level} This phase transforms the model into a flat model of processes that are deployed in a topology of cloud resources.
\item {\bf Monitoring}. The last step  collects  monitoring data from all used resources and applications.  
 Collected data and developed models  can help  identify  performance anomalies, and provide support  
 to the  autonomic principles of a Platform as a Service.  The primary aim is to reduce human intervention, cost, and the perceived complexity by enabling the autonomic platform to self-manage applications~\cite{Tolosana-Calasanz:2015mw}.
\end{itemize}

\textcolor{black}{The smart building scenario discussed at Section~\ref{sec:mod} was developed by means of the Spark model of computation,
thereby the problem is solved in functional terms exclusively. In our approach, the functional level also contains the required resources but in an abstract way. For instance, each EnergyPlus job can be modeled with an associated resource, indicating the maximum degree of parallelism. Then, in the operational level, the actual resources from a resource pool could be incorporated. If this number is less than the number of jobs, then the maximum degree of parallelism could be accomplished. The advantage of doing so is that we can now simulate what-if scenarios (e.g. by varying  workload prediction characteristics, such as arrival rate, predicted execution time, etc.), and study how the execution of EnergyPlus jobs arriving at time $t$ with a number of resources can affect the performance of jobs arriving at time $t + 1$. As a result the analysis, the implementation could follow a policy thereby resources are used in a more efficient way in comparison to Spark's approach.}

On the other hand, the semantics of the component-based language will be defined formally in terms of ordinary PNs~\cite{Murata:1989nt} in order to translate all the advantages derived from a mathematically based model to the methodology  --e.g. Analysis, Verification, or Equivalence Relations. 
The consideration of PNs is based on the natural descriptive power of concurrency, but also on the availability of analytic tools coming from the domain of Mathematical Programming and Graph Theory.  Moreover, taking into account that PNs are executable specifications, PN models can also be simulated.  In summary, PNs are a formalism that combine simulation and analysis techniques. They can be exploited in order to improve the understanding of the system, and to obtain performance or economic boundaries. A survey of PN tools can be found in~\cite{Thong:2015lt}.

The final Petri Net models can be used as a universal specification across multiple platforms and languages. Figure~\ref{Figure0}  shows how the PN tool box can be exploited at all the methodological steps.  The translation of the models specified by a Petri net to an actual system with the same behavior, or the development of efficient interpreters  that directly execute  the models have received considerable attention from researchers over the years~\cite{Basile:2007aa,Moreno:2008aa}. Regarding the last step, we also described our autonomic-based architectural approach in~\cite{Tolosana-Calasanz:2015mw}.

\subsection{Methodological Phases Addressed in this Paper}

As we have just seen, our methodology consists of a sequence of phases.
This paper focuses on the first two ones, where the functional and operational levels are developed, but we also highlight how the Cloud infrastructures can benefit from our approach.

Our models for CDFAs can be exploited to understand the most appropriate approach for solving the problem: Strategy(ies) for decomposing the problem into processes, communication needs and resources required for satisfying
functional and non-functional requirements.

The proposed approach decomposes the construction of the model in phases:
\begin{itemize}
  \item \emph{Functional level}.  Parallel programs solve big problems by simultaneously executing different parts of the problems on different processing resources, which is possible if the problem contains exploitable concurrency, and the first design phase is {\em Finding concurrency}  \cite{Mattson:2004cs}.  This level can be itself divided at least into two different levels:
 \begin{itemize}
 \item \emph{Control level}. At this phase, the designer perceives a CDFA as a set of computational threads, following predefined computational processes. Such computational threads can request, in a competitive way, different quantities of a finite number of shared conservative resources, and they can interact throughout the production/consume of dataflows. The attention is focussed on the study of the problems arising when the shared resources must be granted to a set of concurrent and communicating processes.
 \item \emph{Data-flow level}. The data-flow level may be the result of a refinement of the control level incorporating data and functions to the model.   The result of a data-flow can be part of the control flow represented by conditions or guards, and the control level triggers the execution of functions and data transfers.
 \end{itemize}
  \item \emph{Operational level}. The model is transformed to take into account the execution environment, adapting the functionality to the set of available resources and physical channels used to routing dataflows. At this phase, the actual characteristics of the resources used arise, such as resource capacity, economic cost, or performance.  The aim of the operational level is to refine the functional design and move it closer to the involved resources. This level may be again divided at different levels, the most noteworthy of which are as follows:
  \begin{itemize}
  \item \emph{Enrichment of the model with target platform features}. At this phase, the model is enriched with specific information of the target platform such as execution time, cost or performance variability.  We can analyse at this phase how expansive (cost/time) is  to compute a function or to share information. We can specify a number of what-if scenarios such as if there is hardware support for a shared memory, or if nodes are connected by slow or fast connections.
   \item \emph{Mapping to the resources}. The functional level specifies the concurrency in terms of computational processes, data transfers and involved resources.  The next step in this phase is  to specify  how this concurrency can be mapped onto the resources.  Mattson et al.~\cite{Mattson:2004cs} call this phase the {\em  algorithm structure design space}, and identify  three major principles: Organisation by tasks,  data decomposition, and flow of data.  For example, a matrix multiplication can exploit the data structure by decomposing it in chunks that can be operated concurrently, but if the matrix multiplication is done in the context of a problem with data injected in a stream fashion, a pipeline pattern can be more adequate. 
   \item \emph{Modelling of Supporting Structures}. We can immerse in the modelling of details of the supporting specific target frameworks and programming mechanism. For example, the framework can support  master/worker for dynamically balance the work among resources and queues for supporting a large number consumers accessing it.
  \end{itemize}
\end{itemize}

\section{Case Study: the Wavefront Algorithm}
\label{sec:cas}

In order to illustrate our methodology, we are making use of the Matrix-Vector Multiplication problem in streaming fashion, in particular, the Wavefront
Algorithm, which represents a simple solution for large arrays. The wavefront is a paradigmatically, regularly structured abstraction that allows developers to focus on simple sequential programs to create very large parallel programs. In~\cite{Yu:2009ko}, authors present abstractions as {\em All-Pairs},  {\em Wavefront}  and {\em Makeflow} highlighting the idea that due to their regular  structure, they are more tractable to provide robust and predictable implementation of workloads. In this sense, abstractions are similar to {\em systolic arrays} and {\em wavefront arrays} studied by Kung and Leiserson, sets of regularly interconnected simple processor operating respectively in a synchronous or asynchronous way~\cite{Kung:1982kg}. By using the regular structure and declarative specification, a wavefront may be materialised in different ways on distributed, multicore, and distributed multicore systems showing different performances.

The parallel pipelined wavefront algorithm was originally described by Lamport in his classic paper~\cite{Lamport:1974aa} to execute loops on a multiprocessor computer.  Since then, researchers have proposed performance models and solutions for specific applications and platforms: Algorithms for determining the optimum wavefront partition into sections assigned to individual processors~\cite{Sinharoy:1993aa}, models for performance analysis of wavefronts implemented  on message passing environments in large-scale distributed systems \cite{Hoisie:1999th},   large scaled shared memory multiprocessors~\cite{Manjikian:2001aa},  Commodity Processor Cluster Systems \cite{Mudalige:2006aa}, or  distributed clusters of  distributed  of CPUs and GPUs~\cite{Pennycook:2012aa}.   However, previous analytic performance models can not be generalised or reused  in unexplored new generations of high performance computing system such as the cloud~\cite{Lewis:2000aa,Mudalige:2006aa}. The interest in models for performance analysis of wavefronts is due to  the broad spectrum of applications. It can be found  in simulation problems in economics and game theory~\cite{Yu:2009ko}, problems of sequence alignment via dynamic programming in genomics~\cite{Anvik:2002fe,Alves:2003gf,Serfass:2013uk},  real-time video coding~\cite{Tan:2009ao},  stereo vision and obstacle avoidance \cite{Diotalevi:rm}, or optimal motion planning of multiple robots \cite{LaValle:1998gf}.

Beyond its applicability, the election of a wavefront presents interesting implementation and performance modelling challenges on distributed memory machines because they exhibit a subtle balance between processor utilisation and communication cost~\cite{Hoisie:2000vl,Clauss:2008pb,Yu:2009ko}. Due to its regular structure, the wavefront array provides a concise specification, but it also imposes some message passing constraints that can limit the available parallelism in the algorithm, and it is not easy to reason about its behaviour or performance in an intuitive way.

Let us examine how an algorithm from Linear Algebra, a Matrix-Vector
Multiplication problem in streaming fashion,  can be executed on a square, orthogonal $3 \times 3$ wavefront array (Figure~\ref{Figure4}).
\begin{figure}
%\vspace{6cm} \special{picture FPNWF scaled 1000}
 \includegraphics[width=0.8\textwidth]{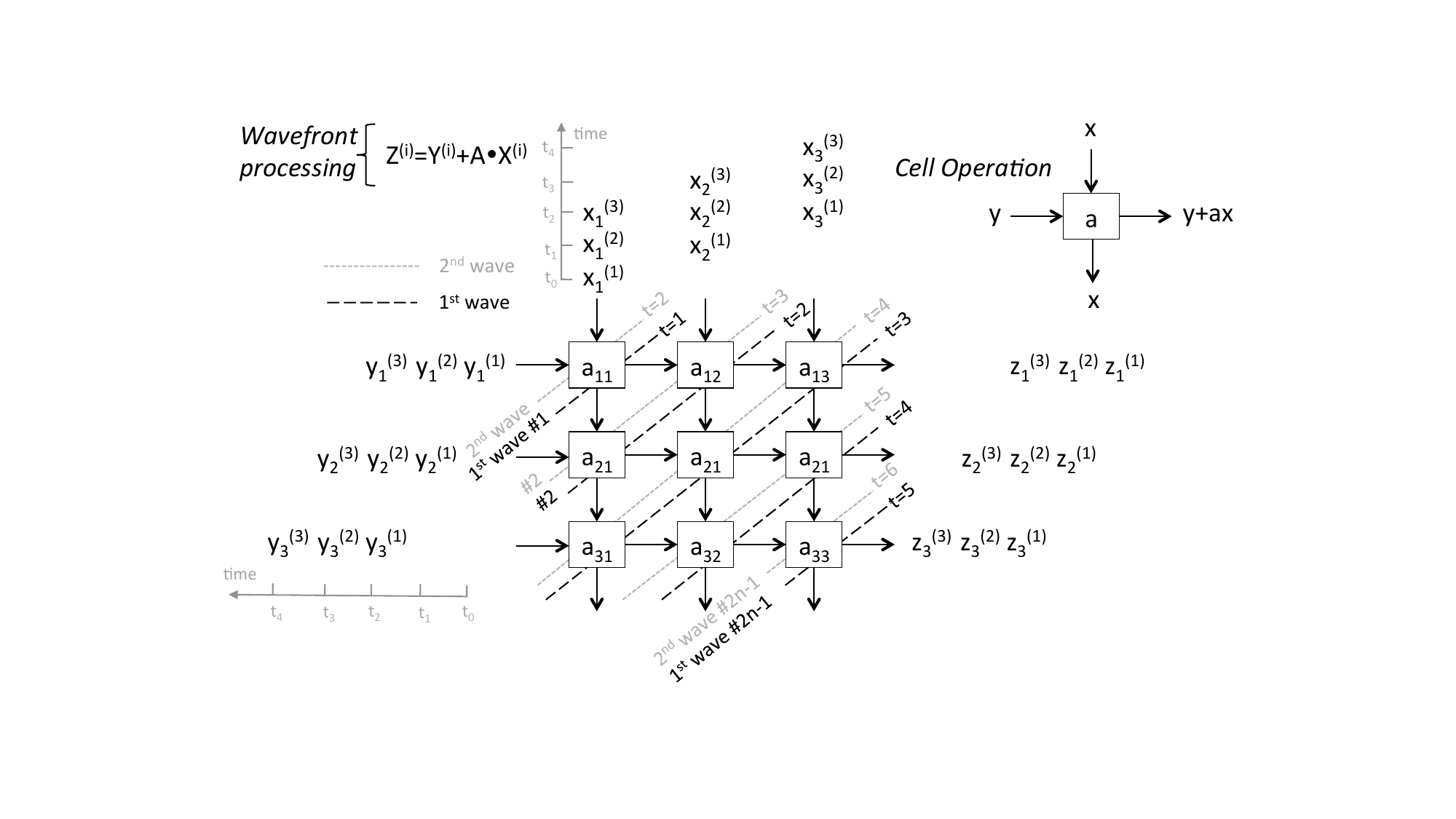}
\centering
\caption{Wavefront processing of $Z^{(k)} = Y^{(k)} + A \cdot X^{(k)}$, $k=1,2,...$.}
\label{Figure4}
\end{figure}

Let $A=(a_{ij})$ be a $3 \times 3$ matrix, and let $X^{(k)} = (x^{(k)}_i)$, $Y^{(k)} = (y^{(k)}_i)$ and $Z^{(k)} = Y^{(k)} + A \cdot X^{(k)} = (z^{(k)}_i)$ be $3 \times 1$ matrices for $k = 1,2,3$. Initially, the elements of $A$ are stored in the array of processors ($a_{ij}$ in processor $P_{ij}$). The elements $x^{(k)}_i$, for $k=1,2,3$ are stored from data streams on the top of the i-th column of processors. The elements
$y^{(k)}_i$, for $k=1,2,3$ are stored from data streams to the left of the i-th
row of processors. The computational process starts with processor $P_{11}$, where  $y^{(k)}_1
+ a_{11} x^{(k)}_1$ is computed. The appropriate data is then propagated to the
neighbour processors $P_{12}$ (the result of $P_{11}$) and $P_{21}$ (the input
data on the top of $P_{11}$, $x^{(k)}_1$), which execute their respective
(similar) operations. The next front of activity will be at processors $P_{31}$,
$P_{22}$ and $P_{13}$. At the end of this step, $P_{13}$ outputs
$z^{(1)}_1$. A computation wavefront that travels down the processor array
appears. Once the wavefront sweeps through all the cells, the first computation
for $k=1$ is finished. Similar computations can be executed concurrently with the
first one by pipelining more wavefronts in succession immediately after the first
one. The wavefronts of two successive computations never intersect, since
once a processor performs its share of operations for a given computation, the
next set of data that it will receive can only be from the next computation.

\section{Modular specification of a CDFA as a Native Cloud application}
\label{sec:spec}
In this section, we discuss about the characteristics of Native Cloud Applications (CNA) and we describe how to build our Petri-net-based functional and operation models from
our methodology.

\subsection{Cloud Native Applications}
Native cloud applications (CNA) are designed to  take full advantage from the cloud computing  characteristics. Fehling et al. identified five essential architectural properties  to  make up a CNA \cite{Fehling:2014aa}, which sets out the steps and patterns that need to be taken into account. The first property is the {\bf decomposition} of the functionality in chunks of distributed functionality.   A CNA is made up of components that can be scaled out independently. Decomposition can be based on layers, processes, or data flows.  Wavefront and dataflow computations are characterized by a data dependent flow of computation \cite{Lewis:2000aa}. When the data flow is regular and static, a pipeline pattern is the best option to represent the function decomposition, and alternatively an event processing network  (EPN) is the the more adequate pattern to represent a data flow with irregular interactions and unpredictable intervals \cite{Mattson:2004cs, Etzion:2010in}. Independently of regular or irregular features,  data flows are made up of two type of constructive elements that can be found in  popular  stream processing engines designed for the cloud  such as Flume, Storm, Spark,  or Flink:  Computational processes that provide a certain function that is performed on input data and produces output data, and data transmission processes that support communication among computational processes. 

The next two properties correspond to the design space  or operation model that comprises the analysis of the application {\bf workloads}  and   the analysis of how the application handles {\bf state} and how expensive it is to share information. The analysis of application workloads involves the way a cloud deployment model can provide a pool of resources that can be automatically managed to provide elasticity to react to varying workloads, and mechanisms to  enable a pay-per-use billing.  {\em Application profiling} is strongly associated with the workload analysis. Profiling must collect a  large amount of data generated by the cloud resources and  forecasting models are fed with these  data to analyse  resource contention and service degradation. A survey on forecasting and profiling models for cloud applications can be found in  \cite{Weingartner:2015aa}.   An important aspect related to the workload is to consider the  dynamic  nature of the cloud, which can be caused by   performance variation of machine instances  offering the same capability, and   by services that are deployed, updated and destroyed all the time  giving rise to a dynamic competition for shared resources \cite{Yeo:2011cy, OLoughlin:2014ww}. CDFAs on the cloud are  complex systems   where customers and resources have not identical characteristics , and exponential distribution does not adequately model observed inter-arrival and service times. Therefore, traditional queuing systems  are not feasible as forecasting models to obtain  accurate performance evaluations. In these cases, we can use approximate methods to compute performance bounds, and  complemented with simulation \cite{Harrison:1992vn,Campos:1992kh,Khazaei:2012rg}. In addition to the mean service time and mean inter-arrival time, the  coefficient of variation of  resources and inter-arrival time  has been proposed to  introduce the dynamic nature of cloud applications and streaming applications on the cloud  \cite{Yeo:2011cy,Schad:2010ij, Khazaei:2012rg,Lohrmann:2015rt}. In our case, profiling data is essential to feed our models with time distribution annotations to estimate bounds. Besides, the explicit modeling of resources at different levels of abstractions supports the subsequent analysis of resource contention by simulation. As regard {\bf state analysis},   in streaming applications,  pipeline tasks or stages act as data transformation activities and can store or not the state between different executions. With elasticity in mind, it is important to specify which components of the application are stateful and which are stateless. In our wavefront case study, all components are stateless.

Finally, the two last properties are {\bf Component refinement} and {\bf Management components} to support  {\em elasticity} and {\em resilience}.  Solutions for the latter are presented by  patterns such as load balancers, or elastic queues.  Component refinement models can be based on middleware such as message-oriented middleware to  support loose coupling,  schedulers that run tasks on opportunistic resources or prioritise QoS enforcements, or multi-tenancy solutions to share resources and components. A catalog of patterns for implementation mechanisms can be found in \cite{Mattson:2004cs,Fehling:2014aa}.  Additionally, elasticity and resilience are also intrinsically linked with resource management, scheduling and autonomic principles with direct effect on performance and cost~\cite{Marinescu:2013ty}.  Although this paper focuses on the first four properties, the utility of formal models for the modelling of 
 components supporting the two last properties is shown in  previous works:  In \cite{Tolosana-Calasanz:2012oy,Tolosana-Calasanz:2011ig,Tolosana-Calasanz:2015mw,Tolosana-Calasanz:2016fk,Medel:2016aa} are presented specifications of strategies on cloud  for resource management following autonomic principles  at the application level for streaming and scientific workflows.  It is also important to highlight the work of  Brogi et al.  \cite{Brogi:2015aa} that  propose to  extend TOSCA with the use of Petri nets for modelling management operations  of complex applications over heterogeneous clouds.

\subsection{Modular construction of the functional Petri Net model of a CDFA}

The construction of the functional model is based on the identification of the basic modules that compose an application of this class. These modules are the Computational Processes (CPs) and the Data Transmission Processes (DTPs). CPs accomplish functional operations and transformations
on data, and DTPs allow data dependencies to be conducted among CPs. Both CPs and DTPs need resources to accomplish the corresponding operation, and these resources also appear in the model, but at a conceptual, and generic way. Later, in subsequent model refinements, specific
resource constraints of different characteristics be added, such as limitations in parallelism, capacity, or economic cost. %R1. Remove etc.

\begin{itemize}
\item \textbf{{\em Characterisation of a Computational Process}}. A CP can be viewed as a type of elementary computational task of an application to be applied to a set of data elements coming from different input data streams. We assume that a CP can consist of multiple instances, called Computational Threads (CT), that are executed concurrently. By analogy with a programming language, we could see a CP as a concurrent program which consists of multiple processes (threads).

\begin{figure}
%\vspace{5cm} \special{picture FPNWF scaled 1000}
\includegraphics[width=0.8\textwidth]{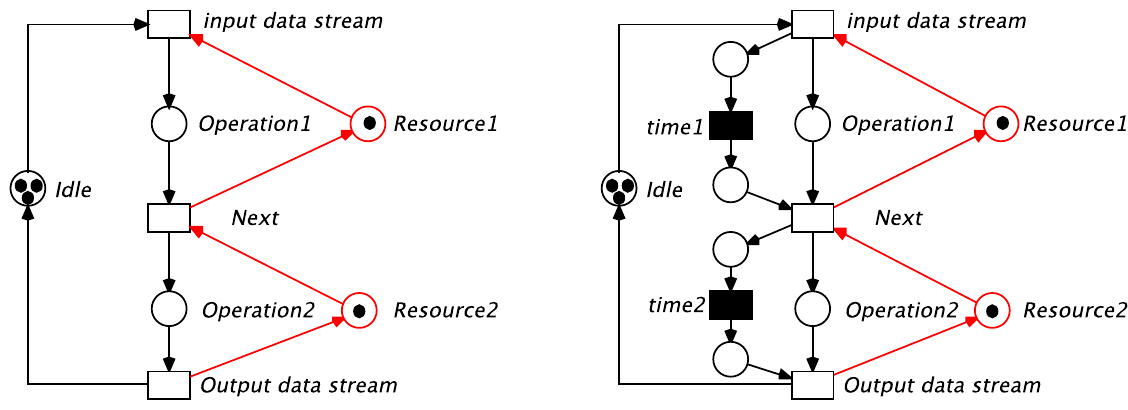}
\centering
\caption{A Computational Process with Resources composed by 2 sequential states, each one requiring a different type of resource. a) Untimed model; b) Timed model assigning $time_i$ units of time to the execution of the $operation_i$.}
\label{Figure1}
\end{figure}

A CP is modelled as a Petri Net, $\N$, and a CT as a token that moves through $\N$. The places (partial states) of $\N$ are related to the different operations (either transformations, handling or assembly/disassembly operations) to be carried out by the thread (See~Figure\ref{Figure1}). There is a special place named \emph{Idle} representing the inactive state of the threads and its initial marking is the maximum number of supported threads executing simultaneously this CP. Each transition of $\N$ represents a state (e.g. task) of a process. Therefore, when the flow of the process makes progress through the transitions, the final state can be reached --representing the end of the computation for the input data elements, the production of the output data elements and the restarting of the thread for the processing of the next data elements on the stream. A CP has distinguished input points (output transitions of the  \emph{Idle}  place) of data elements from the input streams and output points (input transitions of the \emph{Idle}  place) of data elements of the output streams. The execution of a CP is achieved by the execution of a computation path, and several of them can exist in the same CP. A computation path is a sequence of transitions fireable in $\N$, whose occurrence represents the processing steps for a computed data record.

In Figure~\ref{Figure1}.a, a CP with two sequential states is represented. A token in the place \emph{Operation1} or \emph{Operation2} represents a thread executing the code corresponding to the operation 1 or 2, respectively, required by the computational task modelled by means of this CP. A thread executes these operations sequentially following the firing sequence: (1) \emph{Input Data Stream} representing the acquisition of the data records from the input stream to realise the computational task; (2) \emph{Next} representing the end of the operation 1 and the beginning of the operation 2; (3) \emph{Output Data Stream} representing the delivery of the data records obtained after the computation to the output stream. The model presented in Figure~\ref{Figure1}.a is untimed. The addition of timed information to a CP is introduced by the addition of a sequence place-transition-place in parallel with a process place representing an operation of the computational task that consumes time. The transition added is labeled with time information representing the duration of the computational operation. In Figure~\ref{Figure1}.b, the CP from Figure~\ref{Figure1}.a is represented by assigning \emph{time1} and \emph{time2} units of time to the execution of the operations 1 and 2, respectively, according to the previously announced construction for the introduction of timing in the model. Observe that all transitions of Figure~\ref{Figure1}.a are immediate that is, do not consume time.

\item \textbf{{\em Characterisation of a Data Transmission Process}}. A CP can transmit data elements to other CP in the form of a stream sent by means of a physical/virtual device such as a FIFO queue implemented in memory or a communication channel in a communication network. That transmission behaviour is captured by a \emph{Data Transmission Process} DTP. A data element to be transmitted is modelled as a token that moves through a Petri Net, $\N$, representing an elementary DTP with capacity for a single data record (see Figure~\ref{Figure2}). The places of $\N$ are related to the states in which a data element can be in the transmission device. The transitions of $\N$ allow a data record to progress from the source to the destination. The construction of a model that represents the transmission of $k$ data elements (sequentially ordered) requires the concatenation of $k$ of these elementary DTPs. The model in Figure~\ref{Figure2}.a is untimed and the firing sequence of transitions \emph{Begin Transmission} and \emph{End Transmission} represents the movement of a single data element of a stream from the source (the final transition of any kind of Process) to a destination (the initial transition of any kind of Process). 
%The addition of timed information to the DTP is introduced by the addition of a sequence place-transition-place in parallel with a the place representing the transmission state of the data record. The transition added is labeled with time information representing the duration of the transmission throughout the communication device.
In Figure~\ref{Figure2}.b, the DTP of Figure~\ref{Figure2}.a is represented by assigning $time_1$ units of time to the transmission of a data element (according to the previously announced construction for the introduction of timing in the model). 
%Observe that all transitions from Figure~\ref{Figure2}.a are immediate, that is, they do not consume time.

\begin{figure}
%\vspace{4cm} \special{picture FPNWF scaled 1000}
\includegraphics[width=0.8\textwidth]{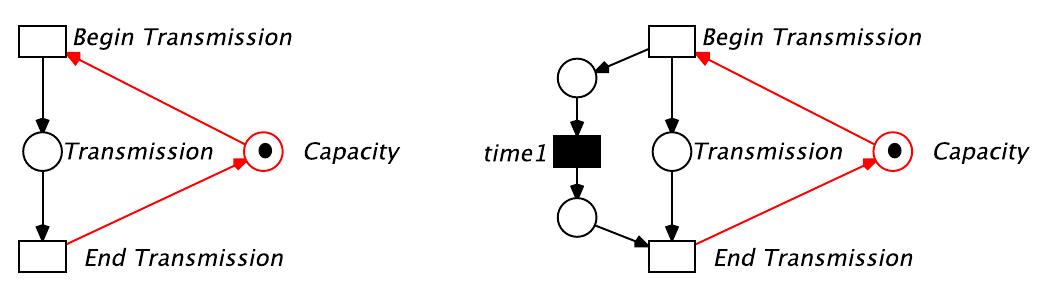}
\centering
\caption{A Data Transmission Process with capacity for a single data record. a) Untimed model; b) Timed model}
\label{Figure2}
\end{figure}

\item \textbf{{\em Incorporation of Resources to each Computational and Data Transmission Process}}. We consider any hardware/software element part of the execution environment (e.g. a processor, a buffer, a server, or a communication channel) as a resource with a given capacity.
In the case of a buffer, its capacity can be the number of positions to allocate elements.  Similarly, a processor may have a number of cores that can be considered as its capacity. Moreover, in the execution environment, there are several resource types and for each of them, a number of identical instances can be available, representing either the number of available copies of the resource to be used (or its capacity). In all cases, the considered resources are conservative, i.e. there is no resource leakage. 
On the other hand, each state of a CP, for its corresponding processing step, requires a (multi-)set of resources (including the buffering capacity to hold the thread itself). In our model, a resource type is represented by means of a place whose initial marking represents either the number of available copies of the resource or its capacity. A resource place has input (output) arcs to (from) those transitions of a CP that moves a Thread to (from) a state that required (used) a number of copies of this resource type. In the case of DTPs, resource places represent the capacity of the storage device for transmission. The CP of Figure~\ref{Figure1}.a requires two different types of resources that are modelled by means of places \emph{Resource1} and \emph{Resource2}. It should be observed that the CP requires a copy of \emph{Resource1} to realise the operation 1 and a copy of \emph{Resource2} to realise the operation 2. In the DTP of Figure~\ref{Figure2}.a, the resource place is the place named \emph{Capacity}, which represents the size of the storage in the transmission device measured in terms of the number of data records. In the figure, it is equal to one (the initial marking of the place \emph{Capacity}).

\item \textbf{{\em Construction of the global model by composition of the Modules with resources}}. In order to obtain the global model of the Streaming application, a number of CPs with their corresponding DTPs (accomplishing the data dependencies among them) must be composed.
Besides, the resources needed must also be considered at this step. The composition is based on the fusion of the resource places representing the same resource type in the different Modules. The initial marking of the resources, after the fusion, is often computed as the maximum of the initial markings of the instances that have been merged. The other composition operation is the fusion of a transition representing the production of data records of an output stream in a module with the transition representing the consumption of data records of an input stream belonging to a different module. Observe that it is possible to connect directly two CP without intermediate DTPs, one of the processes acts as producer of data records and the other as consumer of data records but without any intermediate buffer.
Furthermore, it is also possible to connect directly two DTPs, this represents the construction of a DTP for a Stream of data elements with a capacity of storage equal to two. We can see examples of these fusions in the wavefront example through the rest of the paper, for instance in Figure~\ref{Figure3}.
\end{itemize}

Last but no least, the construction of the model will be done under the principle of economy of details. That is, only those aspects related with the structure of processes of the application and the use of shared resources by the processes will be explicitly represented in the model. These elementary patterns presented here (i.e. computing and transmission) can be combined to form
any data stream pattern, ranging from simple sequential pipelines, to Directed Acyclic Graphs (DAGs), or the complex
Wavefront Array.

\subsection{Modular construction of the operational Petri Net model of a streaming application}

The functional Petri Net model is derived from a specific algorithm that actually processes a number of given data streams. As already seen, it consists of a composition of computational tasks
and the data dependencies among them. In consequence, a \emph{minimal} number of constraints coming from the final execution environment can be taken into account and, in many cases, the functional model is constructed under a number of hypothesis that may not hold
when targeting a specific infrastructure -- i.e.
the resources required in order to reach the maximum degree of parallelism inherent in the model will not be available, or in case there are resources
available, but the economic cost of its usage exceeds the budget. 
Therefore, refining the functional model with the operational submodel aims at introducing specific resource constraints that may alter either economic
cost, performance or even functionality. The alteration of the expected and observed behaviour at the functional model may even induce
changes into the functional model in order to better target a particular execution environment. In other words, the reason for the operational submodel
is to consider explicitly those actual characteristics of a final execution environment, or to compare the response of the application under different deployment scenarios.
In this section, we refine the Functional Petri Net model according to the characteristics of a given execution environment. Nevertheless, there is a huge variety of different characteristics arising from different
execution environments, leading to different requirements and constraints. As a result, there can be many different ways of refining the functional level. Hence, the following is an illustration about how the operational Petri Net model can be constructed from the Functional Petri Net model in two different situations.

The first one corresponds to the case in which the functional Petri Net model has several DTPs that were initially independent, but finally in the operational model they have to be merged together within the same low-level DTP. The actual refinement procedure is depicted in Figure~\ref{Figure3}. There, three independent DTPs, $P_1$, $P_2$ and $P_3$, are displayed that were already present in the Functional Petri Net Model. Nevertheless, the design decision to be taken is that the three Processes must share the same Low-Level Data Transmission Process of capacity 2. The refinement of the model requires the splitting of each place $s_i$ of a DTP $P_i$ in 2 places: (1) $s_{i1}$ represents the request of transmission to the low level; (2) $s_{i2}$ represents the end of the transmission. These two places are connected with a low-level DTP of capacity 2, as depicted in the figure. Observe that in order to recognise the process requesting the transmission, in the low-level DTP a Polling Algorithm to serve the requests has been implemented that is equal to the Polling Algorithm to send the acknowledgements to High-Level DTPs. The other aspect to take into account in the refinement activity is that in case of having timing information for the processes $P_1$, $P_2$ or $P_3$, this information must be removed before the actual refinement; since, after the refinement of the original information, it has no significance. The reason for this is that in the refined model the consumption of time is in the Low-Level DTP.

\begin{figure}
%\vspace{5cm} \special{picture FPNWF scaled 1000}
\includegraphics[width=0.9\textwidth]{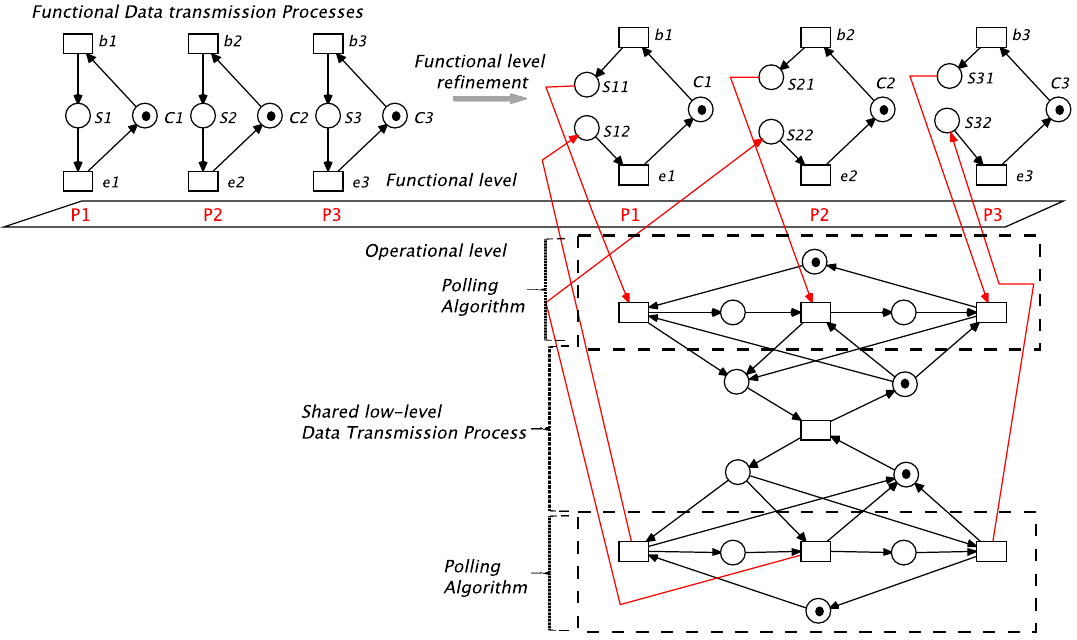}
\centering
\caption{Refinement of the Functional Petri Net model to take into account the operational data transmission process that must be shared for three functional data transmission processes}
\label{Figure3}
\end{figure}

The second case arises if several CPs of the Functional Petri Net Model, which use resource types in isolation, must share the resources between all CPs. This provokes the rise of competitive relationships. 
%In such a case, the refinement of the model is based on the fusion of the resource places of the functional Petri Net model that at the operational level must be shared by the users of the original resource places. The initial marking will be assigned in accordance with the information about the available resources at the operational level. 
A typical scenario for this transformation appears when the number of CPs
is higher than the number of processors and the actual parallelism is
limited.

\section{Functional and Operational models of the pipelined Wavefront}
\label{sec:models}
\subsection{A functional Petri Net model for the Wavefront Algorithm}

\begin{figure}
%\vspace{5cm} \special{picture FPNWF scaled 1000}
\includegraphics[width=0.8\textwidth]{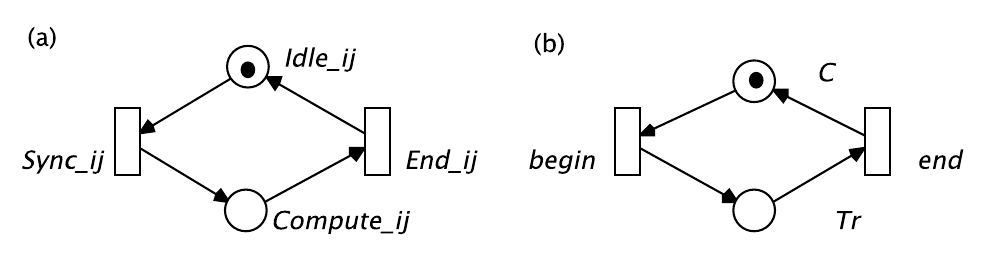}
\centering
\caption{Basic modules for the construction of the functional Petri Net model of the wavefront algorithm: (a) Computational Process associated to a node; (b) Data Transmission Process for the external/internal data streams.}
\label{Figure5}
\end{figure}

The functional Petri Net model of the wavefront algorithm sketched in Figure~\ref{Figure4} is constructed in a
modular fashion. The basic models we need in this case are: (1) A module to describe the Computational Process carried out in a node of the wavefront array; (2) A module to describe the Data Transmission Processes of the input and output data streams to/from the wavefront array. These modules are depicted in Figure~\ref{Figure5}. It should be noticed that they share the same structure with the patterns provided in Figures~\ref{Figure1} a) and \ref{Figure2} a). In order to build the global model, 9 instances of the CP of Figure~\ref{Figure5}.a are needed. The modules of this type belonging to the same row are composed via the fusion of the transition $End\_i1$ with the transition $Sync\_i2$; and the transition $End\_i2$ with the transition $Sync\_i3$. These fusions of transitions represent the transmission of the result elaborated by the column 1 or 2, as input to the columns 2 or 3, respectively, without any intermediate buffering. Each one of the CPs of the first column is composed by a DTP representing the input stream of the corresponding i-th component of the vector $Y^{(k)}$ via the fusion of the transitions $Sync\_i1$ and $end$. Each one of the CPs of the last column is composed by a DTP representing the output stream of the corresponding i-th component of the vector $Z^{(k)}$ via the fusion of the transitions $End\_i3$ and $begin$. Each one of the CPs of the first row is composed by a DTP representing the input stream of the corresponding i-th component of the vector $X^{(k)}$ via the fusion of the transitions $Sync\_1i$ and $end$. Finally, we connect two CPs belonging to the same column but located in rows 1 and 2, or in rows 2 and 3, by means of a Internal DTP describing the flow of the corresponding component of the vector $X^{(k)}$ through the rows of the array. This connection is done by the fusion of the transitions $Sync\_1j$ and one transition $begin$ and the corresponding transition $end$ with the transition $Sync\_2j$ (similarly for the case of rows 2 and 3).

\begin{figure}[t]
%\vspace{7cm} \hspace{1.5cm} \special{picture PNWF scaled 1000}
\includegraphics[width=0.8\textwidth]{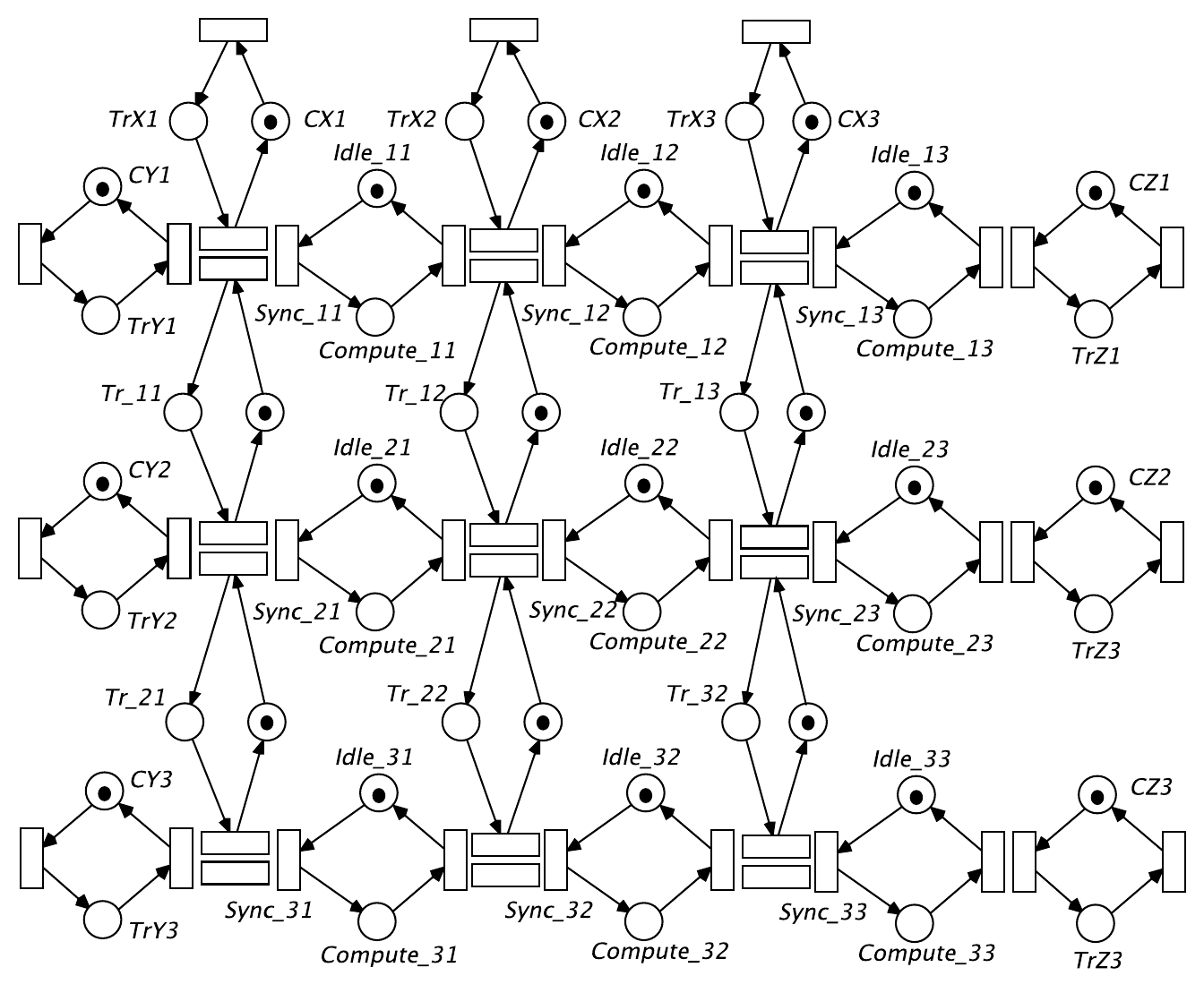}
\centering
\caption{Untimed functional Petri Net model of the wavefront algorithm}
\label{Figure6}
\end{figure}

Figure~\ref{Figure6} depicts an untimed functional Petri Net model of the wavefront algorithm for for $Z^{(k)} = Y^{(k)} + A \cdot X^{(k)}$, $k=1,2,...$. This net model is isomorphous to the flow model in Figure~\ref{Figure4}.

The addition of time to the model of Figure~\ref{Figure6} will be done in the way described in the previous section: adding a sequence place-transition-place in parallel with the place representing the activity that consumes time. The new added transition will be labeled with the time information. In the example, a timed sequence will be added in parallel with each place $Compute\_ij$ representing the duration of the computation accomplished by the CP located at row i-th, column j-th. Moreover, a timed sequence will be added in parallel to each place $Tr$ representing the consumption of time in the transmission of a data element in the corresponding DTP.

\subsection{Operational Petri net Models for the Wavefront Algorithm}

\begin{figure}[t]
%\vspace{7cm} \hspace{1.5cm} \special{picture PNWF scaled 1000}
\includegraphics[width=0.6\textwidth]{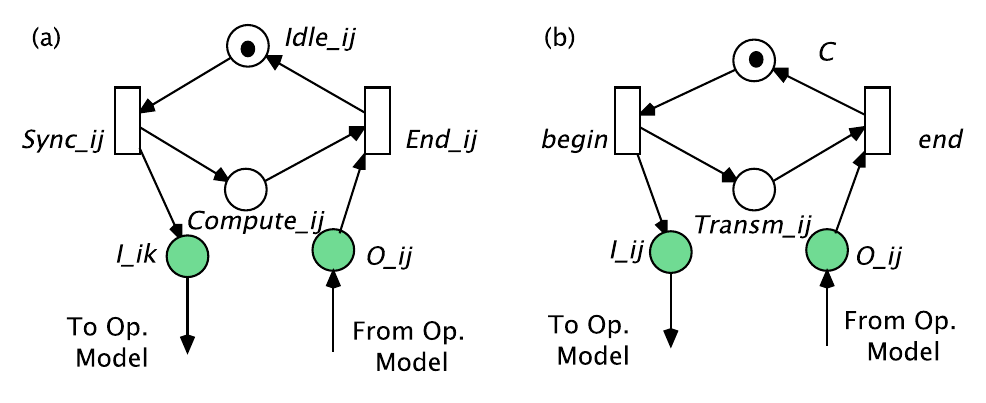}
\centering
\caption{A Strategy for Connecting a Functional Model with an Operational Model}
\label{FigureII}
\end{figure}
The previous functional level can be refined with a number of different operational models with distinct layouts and resources and, hence, achieving different degrees of parallelism. 
In other words, a functional model can also be connected to an operational model in a number of ways. We propose here the strategy depicted in
Figure~\ref{FigureII}, which extends the basic modules for the construction of the wavefront algorithm from Figure~\ref{Figure5}. It can be seen
how both modules (CP and DTP) can be connected to their corresponding computational resource / transmission process in the operational
model, by adding places and arcs as specified in the figure. 
In particular, in Figure~\ref{FigureII} a), if the execution starts, Transition Sync\_ij is fired and the flow then goes to the operational model.
After the operational model conducts the execution, the flow comes back from the operational model to the functional model. The same
mechanism is designed for the DTP module in Figure~\ref{FigureII} b).

\subsubsection{A Grid Operational Model}
Perhaps, one of the operational models that can be proposed in the first place is the grid operational model --based on the
computational processes that appear in the functional model. Therefore, one would need a computational resource for
each CP. Therefore, the operational model needs to be the same in essence as the functional model from Figure~\ref{Figure6}: a 3x3 matrix
of CPs connected with the right and bottom neighbour with a DTP.

The wavefront functional model can be connected to the grid operational model by adding to the functional model the places, transitions and
arcs as in Figure~\ref{FigureII} and then, in particular, by following this specific mapping: 
For each CP and DTP in the functional model, each Place I\_ij will be connected to the underlying transition
Sync\_ij of the operational model, whereas the underlying transition End\_ij of the underlying model will be connected to Place O\_ij. We do not provide a graphical representation model of both the functional and operational models linked together because we understand it will be difficult to display. For that reason, we developed an abstraction and specification language~\cite{Merino:2015aa} that can be found in the Appendix.

\subsubsection{A Pipeline Operational Model}
An alternative operational model could be derived from the fact that the wavefront array works forming sequences (\emph{waves}) of operations. This characteristic was represented in Figure~\ref{Figure4}, thereby the first diagonal (left to right direction) starts an operation which is subsequently continued by the rest of diagonals. In this case, the functional model can be mapped into a sequence (pipeline) of resources, as depicted in Figure~\ref{FigureI}. Therefore, each diagonal from the functional
model is mapped onto a computational resource in the operational model. As there are 5 diagonals in the functional model matrix, a 5 sequential pipeline of computational resources was arranged. It should be noted that each diagonal consists of a
cluster of CPs --with a variable number of CPs depending on the diagonal. We do not provide a graphical representation model of both the functional and operational models linked together because we understand it will be difficult to display. For that reason, we developed an abstraction and specification language~\cite{Merino:2015aa} that can be found in the Appendix.

\begin{figure}[t]
%\vspace{7cm} \hspace{1.5cm} \special{picture PNWF scaled 1000}
\includegraphics[width=0.8\textwidth]{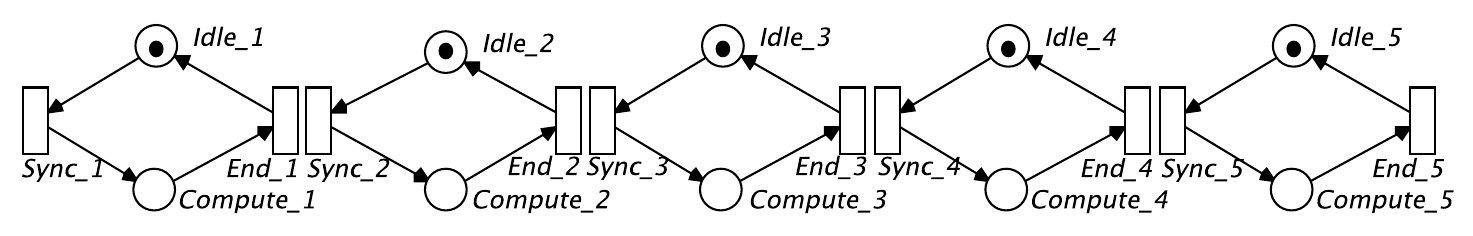}
\centering
\caption{A Pipeline Operational Model}
\label{FigureI}
\end{figure}

Therefore, the connection between the functional and the operational models can be done analogously: 
\begin{itemize}
\item Each Place Compute\_ij of the functional model will be split in two places. One of them will be connected to the underlying transition
Sync\_k, with $k=i+j-1$, of the operational model, whereas the underlying transition End\_k of the underlying model will be connected to the remaining place
of the functional model. It should be noted that the nodes belonging to the same diagonal will be mapped to the same computational resource.
\item Each Place I\_ij of the functional model will be connected to the underlying transition
Sync\_k, with $k=i+j$, of the operational model, whereas the underlying transition End\_k of the underlying model will be connected to Place O\_ij.
\end{itemize}

\subsection{Linking the Models to Real Cloud Resources}
The enormous popularity of the cloud has its origin in the combination of both utility and service computing paradigms. 
As a result, a great number of technologies have been developed in the last years, leading to an overwhelming variety of options of choice, including resources, framework systems, and packaging and tools. In this subsection, we will briefly highlight the type of cloud of resources available and
their particular characteristics that can make any performance prediction challenging.

\subsubsection{Cloud Resource Types}
In general terms, developers can choose among the following cloud resource types:
(i) The conventional physical machines (also known as \emph{bare metal}), (ii) virtual machines that imitate
physical hardware (it contains the operating system and the applications), (iii) containers either comprising the same layers (operating system and application) or only the application, (iv) application packages running on a middleware system (e.g. a web server or application server),
and more recently (v) lambda functions. 

\emph{Virtual machines} (VMs) were one of the first type of cloud resources. A VM can be seen as a piece of software that emulates hardware,  and
typically multiple VMs are executed over the same physical host machine, sharing the same hardware. 
A key component for them is the hypervisor, responsible for conducting the emulation, and they also divide the hardware across the physical machines. 
One of the most important benefits of using VMs is the full isolation they achieve. Nevertheless, VM utilization can sometimes be difficult to achieve, as oftentimes applications to be run do not consume all the resources of a VM. Developers could alternatively try to map multiple applications onto the same VM, but, in such a case, applications would not be isolated. 

\emph{Containers}, on the other hand, represent a way to solve that isolation problem for improving utilisation. A container can be seen as a set of processes where an application is executed in isolation. Multiple containers typically coexist on the same host machine, and each container in it uses the resources that the application on it consumes. Nevertheless, the degree of isolation achieved by VMs is still higher than the one achieved by containers, but containers have much less overhead. The reason for it is that all containers deployed in the same host (physical or virtual) machine share the same OS kernel, and therefore virtualisation is not required. Furthermore, while a VM needs to boot before any application can be executed on it, a container is a group of processes whose execution can be initiated almost immediately, i.e. the overhead is much lower. 
For this reason, containers are rapidly replacing Virtual Machines (VMs) as the compute instance of choice in cloud-based deployments. One of the reasons is the significantly lower overhead of deploying and terminating containers in comparison to VMs. Understanding performance associated with deploying, terminating and maintaining a container is therefore significant.

More recently, \emph{lambda functions} or \emph{serverless-infrastructures} are recently emerging. The main idea behind a function is that they allow developers to execute their code without the need for both provisioning and managing servers. Therefore, developers only load their code into the function infrastructure that is subsequently wrapped as a service, and there is no need to provision for the machines (e.g. switch on a number of VMs or containers). Instead, when triggered by an event, the code is executed and the resource management is completely automated. As being invoked
as a service, functions are closely related to the service oriented paradigm, and to micro-services. Furthermore, functions are often
implemented by container technologies. 

\subsubsection{Cloud Resource Sharing}
As discussed previously, virtualisation technologies and containers, on one hand, allow applications to request the required cloud computational power on demand, as needed, and potentially paying for the use done. On the other hand, from a provider's perspective, they also aim at maximising resource utilisation, and this is achieved by executing multiple tasks simultaneously over the same computational resource. This is appropriate
as tasks are not making use of the required resources at a maximum level over time, but in practice the usage of a resource can vary. Nevertheless,
this sharing of resources in the cloud is subject to resource contention. As a result, performance interference have been reported,
for instance involving virtual machines~\cite{OLoughlin:2014ww,nathuji2010q}, and also
involving container technologies~\cite{medel2016modelling}. Such effects can have a great impact in the execution times of tasks,
in some cases, making the execution time to vary an order of magnitude. Therefore, the methodology discussed in this paper
can be of great interest, as it can analyse the convenience of a deployment configuration depending on the execution circumstances. On the other hand, communication networks can also be virtualised, facilitating the centralisation of control, therefore reducing economic cost and complexity of operating and maintenance.
In the following subsection, we explain how our methodology could be used to integrate real cloud resources.

\subsubsection{Integrating Cloud Resources in our Methodology}
Any of the cloud types described above can be integrated into our model. We propose two ways of capturing the actual cloud resources used and integrate them into our proposed methodology:
\begin{itemize}
\item \emph{Real-time monitoring} of performance execution time and \emph{feeding} of our \emph{operational} models with it. By monitoring these cloud resources (both computational and network resources), real-time performance data can be obtained, and subsequently the operational models can be fed. The addition of time to the obtained model will be done in the way described in the previous sections:
Adding a sequence place-transition-place in parallel with the place representing the activity that consumes time, which will happen in the operational model (rather than in the functional one).
Such a new added transition can be labeled with the time information: A timed sequence will be added in parallel with each place $Compute\_ij$ representing the duration of the computation realised by a CP. Moreover, a timed sequence will be added in parallel to each place $Tr$ representing the consumption of time in the transmission of a data element in the corresponding DTP. As modelling performance interference can be very complex in practice, this option could capture performance unexpected variation in real-time and the models can provide insight on the better alternatives for the mapping of tasks.

\item \emph{Modelling of the actual infrastructure and} integrate it into the models, by refining the operational models. By means of this option,
a model of the cloud infrastructure is developed which serves as a refinement of the operational model. For instance, in~\cite{medel2016modelling},
the authors provide a Petri-net based performance model of Kubernetes~\footnote{http://kubernetes.io/}. Kubernetes is an open source platform that abstracts and automates the deployment of applications across a number of distributed, computational resources. It makes use of container technologies in order to manage and provide computational resources. The model described in~\cite{medel2016modelling} can be annotated and configured with deterministic time, probability distributions, or functions obtained from monitoring data acquired from a Kubernetes deployment. It can be used by an application developer / designer: (i) to evaluate how pods and containers could impact their application performance; (ii) to support capacity planning for application scale-up / scale-down. 
Such models of the infrastructure can be integrated into the functional and operational models analogously by following the same approach in which
the operational model was integrated to the functional model: It was described in Section~\ref{sec:spec}. This approach can capture the whole
behaviour of the computational resources, and therefore it can obtain more opportunities for performance analysis. However, it can require a substantial effort, due to the complexity of the cloud distributed infrastructures.
\end{itemize}

\subsection{Exploiting our Approach in a Cloud Infrastructure}
\label{sec:exploit}
\textcolor{black}{
When using a cloud infrastructure, resources are pooled together in order to process the requests of multiple clients simultaneously. What cloud computing brings as a novelty from traditional data centers is the flexibility in choosing hardware and operating system on demand, as well
as a rapid elasticity. At least in theory, cloud resources promise a rapid automated provisioning to quickly scale out and in. The client, on the
other hand, if using a public cloud just pays for the consumption, as a utility. 
Nevertheless, as already discussed in above in the paper, the sharing and the virtualization technologies are not yet capable to isolate resources
one another and performance interference can arise. This poses some challenges in terms of meeting SLAs, as the amount of the computational
resources that can help meet the SLAs is dependent on the performance interference phenomenon. We believe that our technology can benefit form the cloud elasticity and our obtained models can be exploited in order to mitigate the cloud performance interference problems. In this section, we are providing a way to exploit our methodology based on autonomic principles.
}

\textcolor{black}{
Automated resource provisioning in the cloud often makes use of autonomic principles and exploits the elasticity of the cloud, thereby computational power can be increased / decreased on demand, offering cost-effective solutions. Some of these approaches~\cite{DBLP:journals/tpds/Tolosana-Calasanz17} are often based on the MAPE-K (Monitoring, Analysis, Planning and Execution, and Knowledge) loop. These autonomic systems typically have an objective, related to enforcing QoS as specified in their SLA and which is used as the basis for triggering actions. The loop goes through four phases continuously: (i) During the \emph{monitoring} phase, (near) real-time data is gathered from computational resources of the distributed infrastructure -- this would correspond to our monitoring phase in our methodology, also depicted at Figure~\ref{Figure0}; (ii)
such monitored data is used during this autonomic \emph{analysis} phase in order to determine whether current computational resources are enforcing SLA; in case the objective is not met, (iii) then the planning phase is executed, thereby a (sequence) of action(s) can be triggered, such as adding or reducing the computational power (e.g. this can be accomplished by vertical or horizontal elasticity); finally, (iv) the actions are executed to move current state of the system to a new state where the objective is met.}

\textcolor{black}{The essence for accomplishing this MAPE-K effectively lies in the knowledge (K). Indeed, during the planning phase, the system needs performance models in order to decide which resource mapping option and which resource configuration is more convenient. To this regard, our methodology aims at building performance models and to exploit them with different performance analysis techniques. As we have seen, we can obtain performance minimal / maximal boundaries, and then during the simulation (see Section~\ref{sec:sim}), we can improve the prediction and narrow the gap distance between them, by exploring the combination of parameters, e.g. data income rate and processing throughput. Then, the monitored data, e.g. the actual data income rate and processing throughput, can be used to feed our models and estimate what the CDFA performance will be. The autonomic controller thereafter could decide which is the most convenient mapping solution among the existing ones, and trigger an action that may involve a dynamic change of the current mapping solution. Thus, by means of this mechanism, we can find a relationship between our methodology, the elasticity of the cloud infrastructures, and the phenomenon of performance interference. In our previous work in~\cite{Tolosana-Calasanz:2015mw},
we explored how to exploit these formal-based performance models for the enactment of scientific workflow systems.}  

\section{Performance and Economic Analysis}
\label{sec:analysis}

%Further analysis of concurrency of both models provide interesting insight: (i) In order to extract the maximum degree of parallelism,
%the computational resources where multiple functional tasks are allocated, should be have the appropriate capacity to run such tasks
%in parallel. (ii) Due to the synchronisations among the functional model, diagonals cannot run in parallel: Either even diagonals run in parallel
%or the odd ones. Therefore, with the 5 computational resources allocated, we will have either 2 or 3 computational resources in execution
%at any instant of time, and it can be concluded that resources are under-utilized. A way to solve such a problem would be to modify the initial
%mapping and make use of exactly 3 computational resources.
In the following subsections, we explore the possibilities for both qualitative and quantitative analysis of the functional net model presented in Figure~\ref{Figure6}. Then, we perform an analysis with an enriched operational model (i.e. by adding time inscriptions) that is isomorphous (grid operational model) to the functional model. 

\subsection{Background: Types of Analysis}

The proposed PN model-driven methodology aims at providing  different analysis and prediction techniques that allow developers to assess functional and non-functional properties by means of Qualitative and Quantitative analysis. {\bf Qualitative analysis} aims to detect qualitative properties of concurrent and distributed systems, that is, to decide whether the model  is correct and meets the  given qualitative functional properties (e.g  deadlock freedom).  Qualitative PN analysis can be conducted by means of different techniques: (i) The construction of the state space of the model (reachability analysis) providing a complete knowledge of all its properties -- in case state explosion does not hamper the use of this technique; (ii) Structural techniques in order to reason about some properties of the model, from the structure of the net. 

Among the techniques based on the construction of the state space, standard {\em Model Checking} techniques can be used to explore the state space, which corresponds with the complete set of reachable makings of the PN by the occurrence of transitions. Then, any property to be verified can be expressed in logic terms. If the property is satisfied, then the answer of the Model Checker is just a confirmation of this, but if the property is false, then the model checker gives counterexamples that prove that the property does not hold. The main advantage is that usual properties like deadlock-freeness, home space, maximal sets of concurrently fireable transitions, or mutual exclusions  can be decided. In practice, the applicability of the approach is limited to {\em bounded} systems with a finite state space and with a moderate size. Conclusions hold {\em only for the initial marking} being considered.

The second group exploits the models through the extraction of {\em structural information} from the net. From the structural information, some properties can
be obtained, i.e. the places, the transitions and the token flow relation represented by means of the arcs. 
They can make use of graph theory, linear algebra, convex geometry, or linear programming. Although the use of this technique is not restricted by the state explosion effect, it has a limited decision power (semi-decision algorithms) except for syntactical subclasses of PNs.

From the above comments, we can say that a first phase for exploiting the models is to determine  \emph{the correctness of the adopted solution}, verifying all the good properties expected are satisfied. As a second step, if a property
is not satisfied, the model can be used for \emph{detecting the causes of the problems} or anomalies in it that prevent it from behaving as expected. Then,
the understanding of the causes can lead to a \emph{modification of the model} to correct bad behaviours and to maximise concurrency.

{\bf Quantitative analysis} seeks performance-oriented interpretations of the model such as throughput, utilisation rates, or queue lengths, from which it is possible compute {\em reward functions}. PNs are executable models, therefore extensive simulations can detect errors, which are rare and elusive, and provide us with some performance and reward functions. However, simulations, as an analytical model, cannot  guarantee the absence of errors neither identify under which conditions the simulated performance values will be reproduced.  The most common analytical model used for the derivation of exact  performance measures are stochastic PNs (SPNs)\cite{Bause:2002dq}. The techniques consist on the derivation of performance measures from the reachability graph of the model  from which a Markov Chain is obtained, under certain assumptions on the stochastic specification.  Once again, state explosion can hamper the use of the technique. Additionally, the model assumes exponential distributions for transition delays, which may not be acceptable to model CPU performance, memory speed, sequential and random I/O, or network bandwidth.

\subsection{Qualitative analysis based on Structural Analysis.}
\begin{figure}
\includegraphics[width=0.6\textwidth]{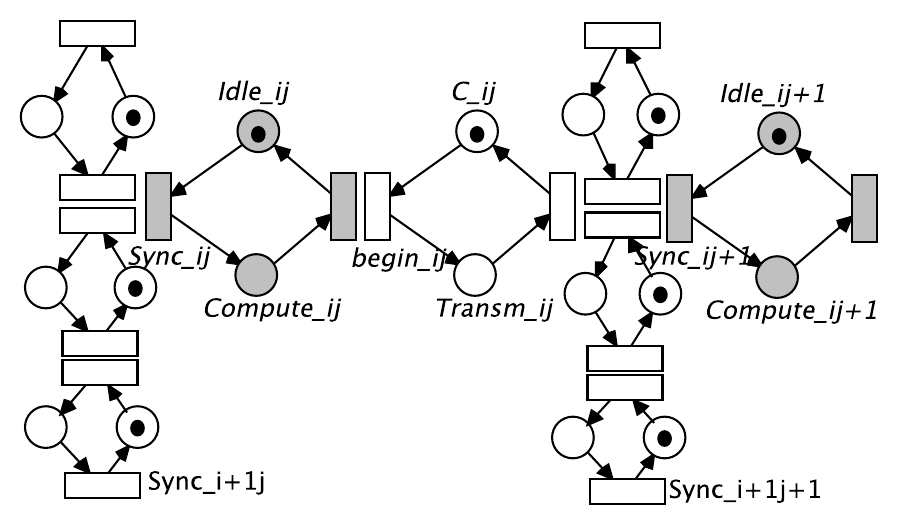}
\centering
\caption{Functional Model of Two adjacent Computational Cells (in dark colour) of the Wavefront Algorithm without Concurrency Limitations}
\label{Fig-mimp}
\end{figure}
First, we determine the correctness of the obtained design at Figure~\ref{Figure6}. This can be realised by exploiting the structural properties of the net  to conclude about behavioural properties. It is a {\em strongly connected marked graph} (a subclass of Petri nets in which each place has only one input and one output transition, being strongly connected in the sense of graph theory). Moreover, all its circuits contain at least one token. From these structural properties and conditions fulfilled by the initial marking, we obtain the following functional/behavioural properties of the model:

\begin{itemize}
\item Any transition of the net is fireable from any reachable state of the net (the net is {\em live}, thus deadlock-free). Moreover, the minimal repetitive sequence of transition firings contains all transitions of the net exactly once (it is guaranteed in live marked graphs from the existence of only one fireable T-invariant equal to vector $\vect{1}$: right anuller of the net incidence matrix). The liveness property is a necessary property of the model to guarantee the execution of a well-behaved wavefront in the array.

\item The wavefronts propagate in an orderly manner, that is, a data element $d_i$ being processed cannot be overtaken by the following data element in the stream $d_{i+1}$. Besides, data element $d_i$, in turn, cannot overtake the processing data element $d_{i-1}$ that started its execution in the previous time slot. This can be concluded after the computation of the maximal difference between firings (in any firing sequence) of a transition $Sync\_ij$ with respect to its: 
\begin{itemize}
\item right neighbor transition $Sync\_i(j+1)$, that is equal to 1. To see this, observe that both transitions are covered by a circuit containing only one token. This circuit enforces a strict alternation in the firing of both transitions starting with the firing of $Sync\_ij$; 
\item left neighbor transition $Sync\_i(j-1)$, that is equal to 0. The reason is the same as in the previous case: the existence of a circuit with a token that enforces the alternation of both transitions starting with the transition $Sync\_i(j-1)$; 
\item bottom neighbor transition $Sync\_(i+1)j$ is equal to 1. Once again, this can be proven by means of the vertical circuit covering both transitions and containing only one token; and 
\item top neighbor transition $Sync\_(i-1)j$ is equal to 0. All these values can be obtained from the so called marking invariants of the net (that in the case of marked graphs are the elementary circuits of the net) and can be computed in a structural way.
\end{itemize}
\noindent This computation can be realised in a Structural Analysis by determining the structural bounds of Synchronic Lead and Synchronic Distance properties between the referred transitions, by means of the corresponding Linear Programming Problems appearing in \cite{Colom-87}.

\item In Figure~\ref{Figure6}, CPs in a \emph{top-right to bottom-left} diagonal can operate concurrently, but CPs belonging to two \emph{top right to bottom left} and \emph{adjacent} diagonals cannot fully operate concurrently. There is a structural limitation in the model, which can be easily proven. Indeed the four transitions $Sync\_ij$, $Sync\_i(j+1)$, $Sync\_(i+1)j$ and $Sync\_(i+1)(j+1)$ are covered by an elementary circuit containing only two tokens. This means that only two transitions out of the four transitions can be concurrently fired. But taking into account the firing relations enumerated in the previous point, only two scenarios are possible: (i) concurrent firing of the transitions $Sync\_ij$ and $Sync\_(i+1)(j+1)$; (ii) concurrent firing of the transitions $Sync\_i(j+1)$ and $Sync\_(i+1)j$. This points out the initial statement about the mutual exclusion in the execution of \emph{top-right to bottom-left} neighbour diagonals.

\item From the previous property, if we extend it to the full array, then we can conclude that we cannot have the nine CPs running concurrently. In contrast, due to the structural limitation discussed before, the maximal concurrency that can be achieved for the model from Figure~\ref{Figure6} is as follows: When the CP associated to Place $Compute_{31}$ (diagonal 1) is in execution, then CPs associated to Places $Compute_{11}$, $Compute_{22}$, and $Compute_{33}$ (diagonal 3) and the CP associated to Place $Compute_{13}$ (diagonal 5) can be in execution concurrently, and the remaining CPs from diagonals 2 and 4 cannot be in execution. On the other hand, when the CPs associated to Places $Compute_{21}$ and $Compute_{32}$ (diagonal 2) are in execution, then the CPs associated to $Compute_{12}$ and $Compute_{23}$ (diagonal 4) can be in execution concurrently, but not the rest of CPs.
\end{itemize}

The previous analysis, without the need for an exhaustive simulation or construction of the state space, points out that it is not possible to have the nine CPs working concurrently. As we discussed, this anomaly or bottleneck limiting concurrency is due to a structural limitation: The existence of the circuits covering four transitions, but containing only two tokens. In order to enable a fully concurrent operation of all the nine CPs, we can decouple any two consecutive CPs in a row by adding a DTP between them.
Such a structural modification can be seen in Figure~\ref{Fig-mimp}, it shows two adjacent computational cells of row $i$ of a Wavefront array: $Cell_{ij}$ and $Cell_{ij+1}$. Both cells are highlighted in dark colour and a DTP module between them. The achieved effect is a decoupling between both cells, so that $Cell_{ij}$ can now start its execution independently from $Cell_{ij+1}$ (Transition $Sync_{ij}$ can now be fired without requiring a direct synchronisation with Transition $Sync_{ij+1}$). 
From a structural analysis point of view, we modified the initial design where we had circuits containing only two tokens, to obtain a structure where 
we enforce circuits with four tokens. In other words, the four transitions $Sync\_ij$, $Sync\_i(j+1)$, $Sync\_(i+1)j$ and $Sync\_(i+1)(j+1)$ from Figure~\ref{Fig-mimp} are now covered by an elementary circuit containing four tokens.

\subsection{Quantitative analysis based on Stochastic PNs}
 \begin{figure}
 \includegraphics[width=0.79\textwidth]{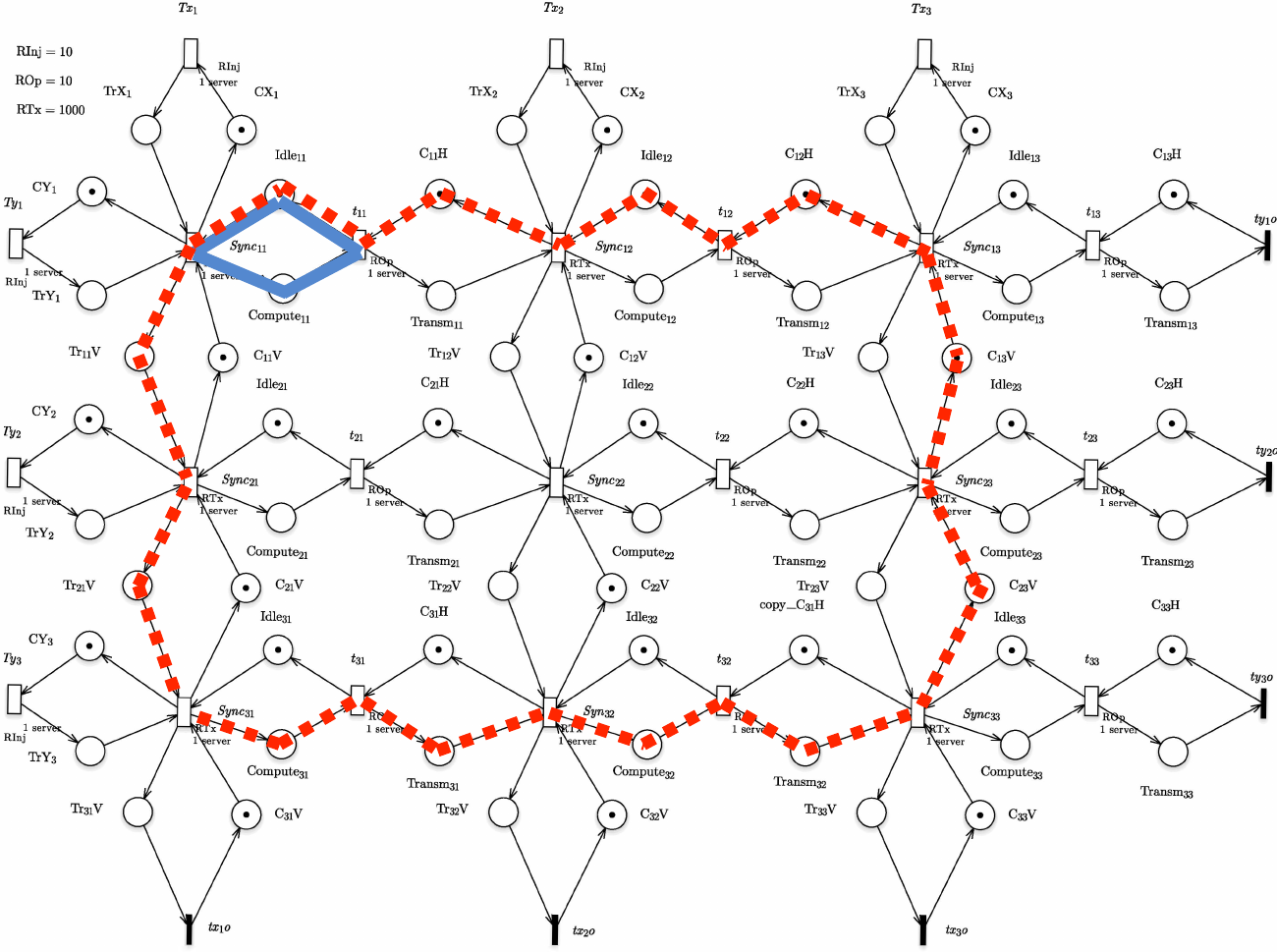}
\centering
\caption{Timed operational GSPN model of the wavefront without concurrency limitations. Blue straight line shows a P-semiflow  that makes maximum the summation of all delays involved in the circuit  divided by the tokens in the circuit. Red broken line shows one of the longest circuits. {\tt RInj=10}, {\tt ROp=10} and {RTx=1000}  represents respectively injection, operation and transmission rates.}
\label{FigureGSPN}
\end{figure}
For the quantitative analysis, we refined the wavefront functional model with the grid-based operational model. In order to perform the analysis, we enriched the model with time. As discussed above, this can be done by adding a sequence place-transition-place in parallel with the places representing the activities that consume time. Such places appear at the operational level, which is built upon the modules from Figure~\ref{FigureII}: The duration of computations (Transitions $Sync\_ij$ from Figure~\ref{FigureII}a), and the consumption of time in the transmission of a data element in the corresponding DTPs (Transitions $begin$ from Figure~\ref{FigureII}b). When the time inscriptions added are a probability density function (PDF), our model is by definition a stochastic PN. Only the use of a negative exponential PDF for the specification of temporal characteristics makes the analysis mathematically tractable. Let us assume that processors service delivery time $(1/\lambda)$  and injection timed transitions $(1/\gamma)$ follow an exponentially distributed random amount of time, with average 100ms (rate =10 data/sec); and a transmission time 100 times faster with average ($1/\beta$=1ms) also following an exponential distribution.

Then, we can introduce this resulting operational PN model directly to GreatSPN2.0.2 (See Figure~\ref{FigureGSPN})\footnote{\tt http://www.di.unito.it/\textasciitilde greatspn}~\cite{Baarir:2009mq}. GreatSPN2.0 (GSPN) is a software package for the modelling, validation, and performance evaluation of distributed systems using Generalised Stochastic Petri Nets and their coloured extension. Recall that, from the previous structural analysis, it can be derived that all transitions will have the same throughput: The minimal repetitive sequence of transition firings contains all transitions of the net exactly once (it is guaranteed from the existence of only one T-invariant: right annuler of the incidence matrix of the net). Therefore, the relative firing frequency vector is {\bf 1}, and we have the same mean cycle time for all transitions. The relative firing frequency vector is a vector that for each component, it contains the mean interfiring time of each transition $t_i$, i.e. the inverse of its throughput.
The tool generates the reachability graph with 1,392,640 states from which a Markov chain is derived. Performance indices like place markings, probability distribution and transition throughputs can be computed. The calculated throughput for all transitions is 3.99 data/sec. The result obtained by the GSPN analysis shows a poor throughput with a performance loss of 60\% for each processor, in comparison with the model that achieves maximum degree of concurrency and all its CPs are executed concurrently.

Furthermore, if we repeat the analysis with a wavefront of dimension 2x2, the throughput for all transitions is 4.69 data/sec with a performance loss of 53\% for each processor in comparison with the model that achieves maximum degree of concurrency and all its CPs are executed concurrently. This is due  to the concurrency limitations found in the structural analysis. All transitions  will have the same throughput, and as a result throughput is determined by the slower transition. Intuitively, the use of a negative exponential PDF, with a high coefficient of variation, makes more likely a slower transition than the expected throughput  with a larger number of processors.

\subsection{Quantitative analysis based on Structural Analysis: Computing performance bounds}

The  use of stochastic PNs for the derivation of exact  performance measures and rewards functions is hampered by two factors: (i) The explosion of the computational complexity of the analysis algorithm and (ii) only the use of an exponential probability distribution function for the specification
of temporal characteristics makes the analysis mathematically tractable.  

In~\cite{Campos:1992kh}, authors present upper and lower bounds on the steady-state performance of marked graphs that can be computed efficiently. To do that, they derive bounds for the throughput of transitions, defined as the average number of firings per time unit (or its inverse, called the
{\em mean cycle time of transitions}, $\Gamma$). From this value, applying Little's Law, it is possible to obtain other average performance estimates
of the model. Under these restrictions they showed results that can be computed in polynomial time on the size of the net model, and that depend only on the mean values and not on the higher moments of the probability distribution functions of the random variables describing the timing of the system. 
Finally, they found that both upper and lower bounds, computed by means of Linear Programming Problems, are tight, in the sense that, for any marked graph, it is possible to define families of stochastic timings such that the steady-state performances of the timed Petri net models are arbitrarily close to either bound.

For the case of throughput upper bounds for strongly connected marked graphs, the computation method is obtained by applying Little's Law to each place of the net. Besides, special marking invariants, derived from the P-semiflows, of the net are used. A P-semiflow is a special sequence of firing of transitions of a Petri net, so that when such firing sequence occurs in it, its marking (i.e. the token distribution) is invariant. Further description about Petri nets and these topics can be found in~\cite{Murata:1989nt}. Furthermore, the dynamic behaviour of Petri nets can be expressed in terms of matrix equations and the following Linear Programming Problem can be derived. Its optimal solution (which can be computed in polynomial time) is a {\em lower bound} for the {\em mean cycle time of transitions} (inverse of the average throughput) that is denoted as $\Gamma^{min}$.

\begin{small} 
\[
\begin{array}{ll@{}ll}
\text{$\Gamma^{min}$  = maximum}  & Y^{T}*\/ Pre \/ * \/  \theta &\\
\text{subject to}&Y^{T}*C=0,  Y^{T}*M_{0}=1,  &\\
                 & Y \geq 0  &
\end{array}
\]
\end{small}

\noindent where $\Gamma^{min}$ is the minimum mean cycle time, $Y^{T}$ is the left annuler of the incidence matrix of the net (P-semiflow), $Pre$ is the pre incidence matrix (i.e. denoting tokens removed by transition firing), and $M_{0}$ is the initial marking.  This means that the  mean cycle times can be computed by the summation of all time delays involved in a circuit (P-semiflow) divided by the tokens in the circuit.  And we obtain the $\Gamma^{min}$ by finding the maximum value of mean cycle times computed by each circuit. In our model, $\Gamma^{min}$ is 101 ms $(1/\lambda)$, i.e., a maximum throughout  of 9,9 data/sec.  Figure~\ref{FigureGSPN} shows one of the possible circuits highlighted by a blue straight line. It  returns the maximum value of mean cycle time  computed by the summation of all time delays.  {\tt RInj=10 data/sec}, {\tt ROp=10 data/sec} and {RTx=1000 data/sec}  represents respectively injection, operation and transmission rates. Therefore,  time delays in the blue circuit is the addition of operation in  {\tt t11} is 1000/10=100 ms, and transmission 1000/1000 = 1 ms giving a total of 101 ms divided by one token in this circuit.

A tight {\em upper bound for the mean cycle time} (its inverse is  a lower bound for the steady-state throughput) is obtained in polynomial time, from the knowledge of the given average service times and the liveness bounds of transitions, which are computed by solving proper linear
programming problems. This bound cannot be improved unless more information from the service times of transitions than their mean values is used.

%On the other hand, the {\bf upper bound} for the mean cycle can be computed by:

\begin{small}
\[ \Gamma^{max}=\sum\limits_{j=1}^{m} \frac{\theta_{j}}{LB(t_{j}) } \]
 \end{small}

\noindent where $\theta_{j}$ denotes the average service time of transition $t_{i}$, and  $LB(t_{j})$ the liveness bound of $t_{j}$, i.e. its maximum degree of concurrency (maximum number of concurrent firings of a transition).  In our example, the maximum liveness bound is 1 for all transitions. Therefore, $\Gamma^{max}$ is given by the addition of $theta_{j}$ corresponding to the longer circuit. The dotted red line in Figure~\ref{FigureGSPN} shows one of the possible longer circuits that defines the lower throughput bound. This circuit comprises four operations (100 ms) and eight transmissions (1ms), and the the maximum degree of concurrency of each transition is one because there is only one token in previous places. More specifically, in our model:  $\Gamma^{max}=408$ ms and a minimum throughput of  2.45 data/sec.

Under the light of this analysis, we can obtain best- and wort-case scenarios in terms of throughput, which is inline with the results obtained from
previous analysis (i.e. the GSPN analysis also reveals that the 2x2 wavefront has better throughput for each output than a 3x3 wavefront).

But the most important result of this analysis is that we can  estimate  upper and lower bounds that specify the space of possible performance solutions (see bounds in Figure~\ref{Figure7}).

\subsection{Quantitative analysis based on Simulation}
\label{sec:sim}

If the obtained stochastic PN model is too large, the generation of all of its states to obtain the underlying Markov chain or even solving
the previous optimization problem may not be feasible.
In consequence, an alternative is to make an analysis based on simulations. Simulations can also be used to explore different what-if
scenarios between the obtained performance bounds. One of its main drawbacks is that it only reports on simulated situations,
since it is not an exhaustive enumeration technique. Besides, it can also be a time-consuming approach, as each of the different settings need
to be simulated and it can be difficult to determine whether the system reaches stability.

Furthermore, one of the challenges for conducting simulations on the model is to choose the appropriate probability distribution functions of response time of processes (i.e. computations and transmissions in our case)\textcolor{black}{, in order to provide with a realistic exploration of all the variability.}
In our previous quantitative analysis based on stochastic PNs, the system throughput is near the minimum bound. Then, the selection of an exponential distribution for task service times is not adequate and results in poor and not realistic performance results.
However, if the service times are not exponential, it can be complex and it relies on
a set of approximations. Such approximations are sensitive to the probability distribution of service times and they even become increasingly inaccurate when the Coefficient of Variation (CoV) increases towards the value of 1~\cite{Yeo:2011cy,Khazaei:2012rg,OLoughlin:2014ww}. 
The CoV is a measure of dispersion of a probability distribution. It can be used to show the extent of variability in relation to the mean of the population. Therefore, CoV is formally defined as $CoV=\frac{\sqrt{Var[X]}}{E[x]}$.
The higher the values of CoV, the higher the dispersion in service times.
Furthermore, the difficulty increases when we try to analyse performance of applications implementing advanced data flow abstractions with a high level of concurrency constraints. 

\begin{figure}

\includegraphics[width=1\textwidth]{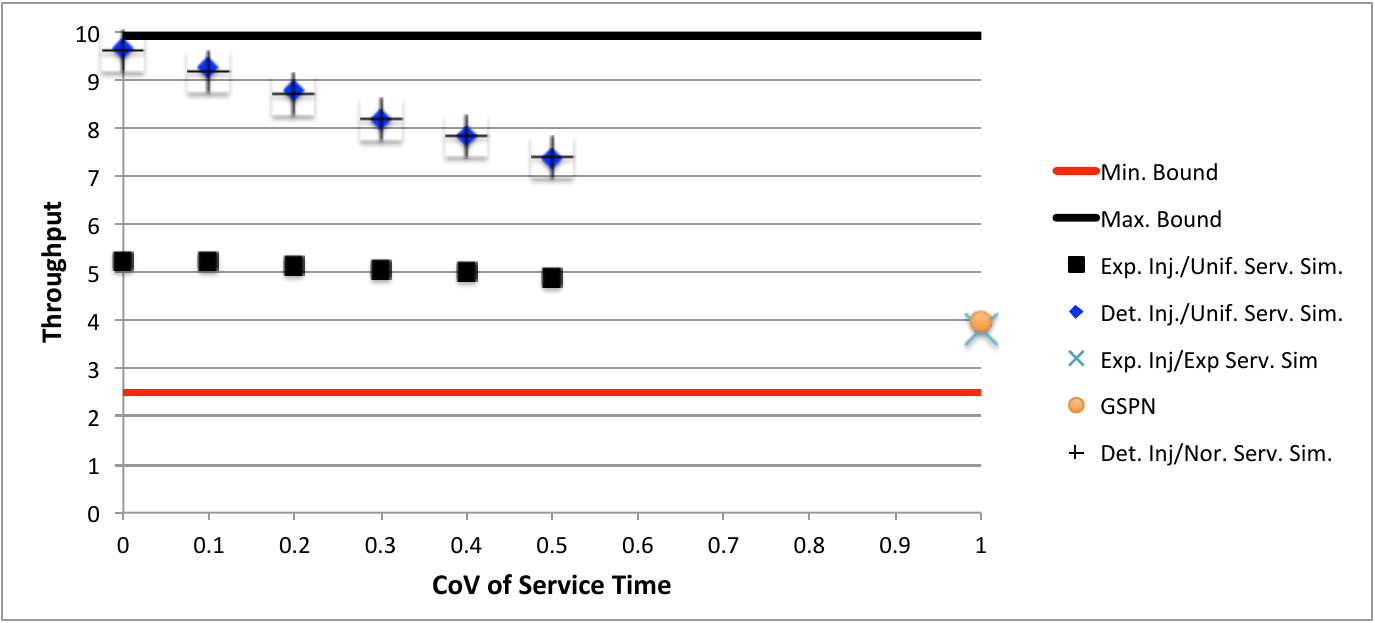}
\centering
\caption{Simulated throughputs with different Coefficient of variation of the service time. }
\label{Figure7}
\end{figure}

Once the throughput bounds for our model are known from the previous analysis, and given the fact that the CoV on service time can have
a high impact on performance, we conducted simulations to evaluate its impact. Such a characteristic can be of great interest, for instance, when
computational resources are subject to unexpected performance variations (e.g. in some clouds).
In order to conduct the simulations, we made use of  the Renew tool. Renew~\footnote{{\tt http://www.renew.de}}~\cite{DBLP:conf/apn/KummerWDSKMRV04} is a Java-based Reference net interpreter and a Reference net graphical modelling tool.

Figure~\ref{Figure7} shows the results of different simulations with mean services and injection times of 10 data/sec.  In order to perform the simulations, we introduced our operational PN model into the {\tt Renew}\footnote{\tt http://www.renew.de}  PN interpreter,
\textcolor{black}{In the example, we considered the transmission time to be negligible compared to the average service time}. The figure  shows the point {\em GSPN} that represents the computed performance by the GSPN tool obtaining the Markov chain, and the  point {\em Exp. Inj./Exp. Serv. Simulation} shows  the same scenario obtained by simulation in Renew, i.e, assuming both service delivery times and inter-arrival times are exponential. The CoV of an exponential distribution is 1. The proximity of these points  shows the accuracy of simulations. Assuming the computing nodes in the cloud are heterogenous and that the performance capabilities of these computing nodes are uniformly distributed~\cite{Yeo:2011cy} between the time of the faster node and the time of the slower node, we conducted different simulations in {\tt Renew}.   Service time CoV ranges in simulations from 0 to 0,5 with uniform and normal distributions. It is not possible to obtain higher values of CoV with these distributions.  {\em Exp.Inj./Unif. Serv. Simulation} shows the impact of CoV on performance assuming an exponential distribution in injections, and uniform distributions of service delivery times. {\em Det.Inj./Unif. Serv. Simulation}  shows the same simulations with a deterministic injection time. Finally, {\em Det.Inj./Nor. Serv. Simulation} shows that simulations with normal distribution of service delivery times provide the same results as a uniform distribution (cross and triangles overlap in Figure~\ref{Figure7}). These results show that the throughput depends on the CoV, but it is independent on the probability density function.

These results also show the relevance of a mechanism to regulate injections rates and avoid bursty behaviours. As it can be seen in Figure~\ref{Figure7}, assuming exponential distribution of inter-arrival times results into performance near 50\%. Without an injection rate regulation of a bursty flow, the service time CoV is less important.  Once the incidence of the inter-arrival time CoV is reduced by any admission control mechanism such as a leaky or token bucket, performance only depends on the Service time CoV improving performance. In~\cite{Tolosana-Calasanz:2016fk}, a mechanism based on token bucket to regulate bursty streams on multi-tenant cloud environments is proposed.

\begin{figure}
\includegraphics[width=1\textwidth]{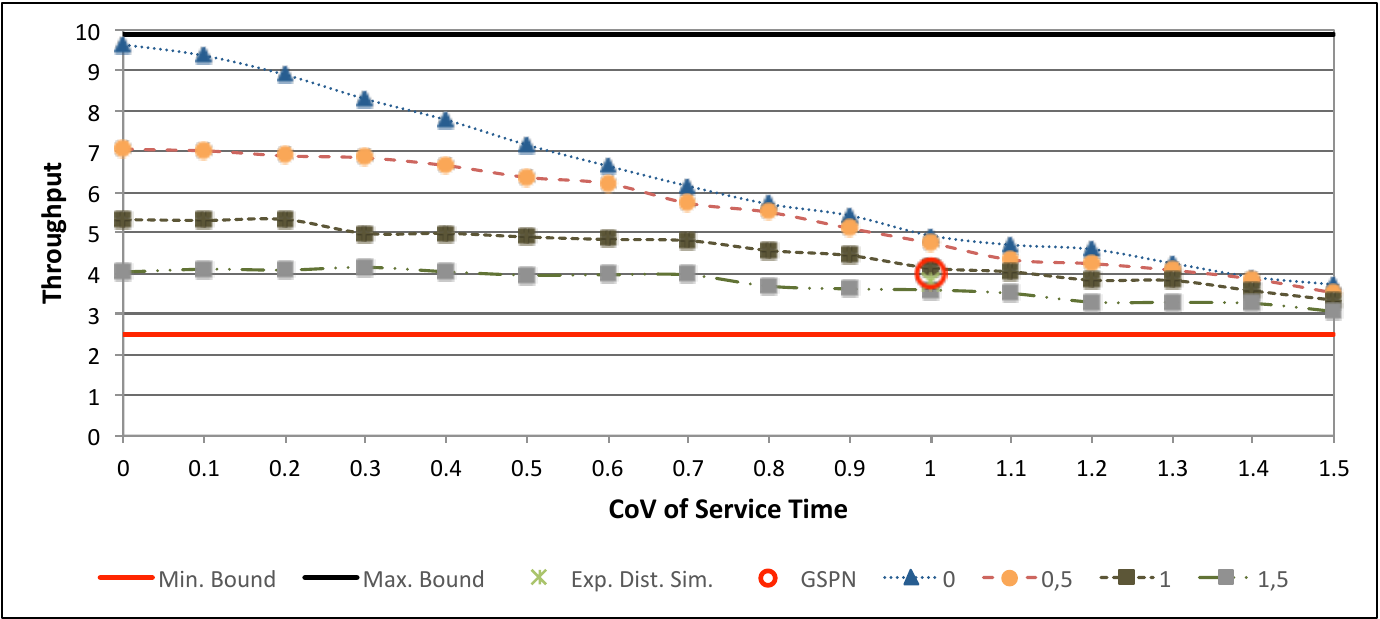}
\centering
\caption{Simulated throughputs with different CoV of  service time and injection rates following a Gamma distribution. }
\label{Figure7b}
\end{figure}

In Figure~\ref{Figure7}, we make use of normal, uniform and exponential probability distributions for inter-arrival rates and processing times, whereas in Figure~\ref{Figure7b}, we make use of a gamma distribution. Figure~\ref{Figure7b} aims to highlight the influence of CoV in the distribution of inter-arrival times and service time using a gamma distribution.  Each point  shows the mean throughput for four simulations  using a gamma distribution with CoV of service time ranging from 0 to 1.5. The figure shows the results for deterministic injection (CoV=0) and CoV=0,5, CoV=1 and CoV=1.5 of injection rate. In this way, we show how simulations are conducted to cover the space of solutions between the minimum and maximum estimated bounds. The purpose is to show that performance does not depend on the probability distribution, but on the Coefficient of Variation (CoV). Besides, with normal and uniform distributions, values of CoV higher than 0.5 cannot be generated. This is why we also used a gamma distribution, which allows us to generate CoV values higher than 0.5.

\subsection{Economic Analysis}
The knowledge obtained from all the previous analysis techniques can be subsequently used to study the economic cost of executing an application.
In this subsection, we also make use of our recurrent example throughout the paper: The wavefront algorithm. We are taking into account the pricing models in~\cite{Gohad:2013fk}, which include the pricing cost associated to different data transfers, CPU time or storage usage.
For the sake of simplicity, we are just considering an economic cost per CPU usage through time,
information that can be added into the model in the following way:

\begin{itemize}
\item Let us assume, under a deterministic timing interpretation, a time $\alpha$ for the timed transitions added to represent the duration of the input of data elements from the streams corresponding to the vectors $Y^{(k)}$ and $X^{(k)}$ (timed transitions in parallel with the places $TrYi$ and $TrXi$, a time $\beta$ for transitions corresponding to the execution of the code of the CPs (timed transitions added in parallel with the places $Comp\_ij$), and a time $\gamma$ for transitions corresponding to the internal transmissions in the wavefront array. A reachable (exact) bound of the throughput of the system can be computed through presented structural techniques~\cite{Campos:1992kh}, obtaining a value equal to the inverse of $max\{\alpha,\beta,\gamma\}$. This is the first relevant aspect to be taken into account from the economical point of view from our model: The cost derived from the CPU time consumed depends on the slower resource (CPU).  Additionaly, under a stochastic timing interpretation, the computed bounds following~\cite{Campos:1992kh}  allow us to derive economical cost bounds.  These bounds can be eventually improved (in~\cite{Campos:1993ir} a search for embedded queueing networks was considered).
   
\item We can do the following economic analysis on the functional Petri Net model. A bound of the mean cycle time for this net (the elapsed time between two consecutive firings of a transition) is the inverse of the throughput, from this value, we can compute the economic cost for the processing of streams of length $k=n$, assuming the cost of the time unit per CPU, $p$: $Cost_{functional} = max\{\alpha,\beta,\gamma\} * n * p * 9$, where $\alpha, \beta, \gamma, n$ represent the average service time, transmission time, injection rate, inter-arrival time, and the wavefront matrix dimension, respectively. It allows the designer to have an accurate estimation of the cost taking into account the pricing applied to CPU time consumed. Note that this analysis corresponds to the functional level and no information on the operational level has been considered. In this case, the model is designed with nine CPs having each nine computational resources in isolation.

%\item[3)] Given the fact that the model in Figure~\ref{Figure6} is a marked graph, an optimal scheduling policy is just the earliest-firing-time policy (i.e. fire the transitions as soon as possible).

\item  In case we wanted to introduce a particular operational level, we would
have to proceed with the refinement of the previous functional Petri Net model and with a similar analysis to the previous one. Besides, in such a refined model, the relationship between economic cost and performance can be studied: i.e.
how performance varies when the economic cost is reduced. Taking into account the quantitative analysis and the simulations,
we can also observe the impact that service time and injection rates variance have in performance. The CoV of service time has been pointed as an additional difficulty in  estimating the number of resources for optimising cost from the client's point of view, or energy and cost from the cloud provider point of view \cite{Yeo:2011cy}. The operational cost becomes a function of the service time and injection rate variances: $Cost_{operational} = f(CoV_{Service Time}, CoV_{Injection Rate})*Cost_{functional}$ . Function $f$ denotes that the cost depends on the CoV coefficients for injection and processing rates, whereas $Cost_{functional}$ is described in the previous point. Both $f$ and the functional cost are application dependent and can be obtained by conducting the analysis performed in this section.

\item The results of previous analysis can be used to reduce the cost in different ways. By knowing the impact of service time and injection rate variances, the performance engineer can integrate and tune a data admission and control mechanism in order to control the performance (as accomplished in \cite{Tolosana-Calasanz:2016fk}). Clients can specify in fixed SLAs the variance of provisioned services and pay for them according to this specification, or at least a way  to run meaningful performance experiments to know it~\cite{Schad:2010ij}. 

\end{itemize}

\section{Related Work}
\label{sec:rel}

There is several related work related with the modelling and  performance evaluation of parallel programs. A thorough revision of all involved aspects  is out of the scope of this paper. As a brief summary,  from the modelling point of view, the approaches  of I.T. Foster \cite{Foster:1995fk} and T.G. Mattson et al. \cite{Mattson:2004cs} are the main guidelines in conjunction with a set  of parallel patterns \cite{C.Pautasso:2006ss,Yu:2010tl}.  More specific to the cloud, Fehling et al. present in~\cite{Fehling:2014aa} the essential patterns for building cloud applications.  Besides, there are also works proposing cloud business models, providing reference models  for value chains in the Cloud \cite{Mohammed2009}.  In this sense, our proposal is compatible with previous methodologies, providing them with a component-based approach and capabilities to reason with the models.

 On the other hand, performance models of computer and communications systems have been studied from many years, basically using  queueing networks \cite{Harrison:1992vn}.  However, the introduction of synchronisation constraints naturally turned the focus to Petri nets \cite{Marsan:1995un}. Performance models of cloud computing have attracted considerable research attention, but likely due to the complexity involved, rigorous analytical approaches have only been adopted by works that focus at the server farm level \cite{Khazaei:2012rg}.  Other approaches for performance analysis try to derive Petri net performance models from UML diagrams with standardised annotations in the  "UML Profile for  Modelling and Analysis of Real-Time Embedded systems" (MARTE)  \cite{Woodside:2014qe}. Unlike these approaches, our emphasis in here is to exploit the inherent nature of PNs for modelling concurrency, and make use of it for modelling the specific aspects of Streaming applications, which involves concurrent computations and data transmissions, and third-party resources.

From the application point of view, for example, we can find some work focusing on performance of the wavefront pattern~\cite{Hoisie:1999rr,Lewis:2000aa,Clauss:2008ph,Yu:2010tl,Dios:2010fv}.  However, these works are not based on formal models, and the same happens with other patterns \cite{Jiang:2010dn,Yu:2010tl}. We can also find some work dealing with scientific workflows and data flows modeled as  Directed Acyclic Graphs (DAG) on grid, cluster and cloud~\cite{Hoffa:2008oe}. In general terms in pipelined workflows, performance is typically measured in terms of throughput, and therefore the throughput is conditioned by the slowest task. For such a reason, it is important that all the tasks execute in the same time, which is challenging due to the variability and heterogeneity of computational resources as well as the programs that execute the tasks. Thus, task merging and workflow transformation is essential prior to mapping the tasks onto distributed resources in order to achieve the minimal variation in execution time of tasks. We believe that our proposal in this paper can help analyze
the influence of the different design decisions in the task-mapping process on workflow throughput and other properties such as economic cost. Other proposals like~\cite{Tolosana-Calasanz:2008ty} utilised Petri Nets for predicting the execution time (makespan) of Taverna workflows at a functional level. 
%Regarding the problem of workflow mapping of tasks and task clustering has been studied deeply in the Pegasus workflow system~\cite{Lee:2008sy} for  DAGs. In particular, in~\cite{Chen:2013fk}, the authors highlight the problems that arise with current task clustering techniques as they are based on over simplified workflow models. They investigate the causes and propose a number of task balancing methods to address these imbalance problems. 
It is worth noting the work for bottleneck detection and performance analysis for data intensive applications, related to {\em Nephele}, the stream engine of {\em Flink} \cite{Battre:2010rr, Lohrmann:2014jt, Lohrmann:2015rt} . However, as these authors point out, the extensive amount of related literature focus only on different abstraction levels.   

Finally, framework systems based on specific patterns, or stream processing engines  such as {\em Flink}, {\em Storm}, or {\em Spark}, Yahoo's S4~\cite{Neumeyer:2010qf}, or IBM InfoSphere Streams~\cite{Biem:2010yq} provide
streaming programming abstractions to build and deploy tasks as distributed applications at scale for commodity clusters and clouds.

\section{Conclusion and Future Work}
\label{sec:con}
 
In this paper, we described a Petri net-based, model-driven and stepwise refinement methodological approach for CDFA  executed over cloud resources. Current available technological solutions (such as Apache Big Data Stack or the solutions by most vendors) provide high-level abstractions that cover different functional and non-functional aspects, but such approaches require from engineers to elaborate every detail of their implementations. Our methodology follows an iterative and cyclic approach, starting from the specification of functional algorithms (specified in the functional models). Then, it continues with the specification of the computational resources available (in the operational models), providing complementary views: Control flow, data-flow, and resources. Such separation of views also enables functional and operational model reuse, and the exploration of what-if scenarios that enable the analysis of the application. Such analysis provides a  quantitative and qualitative evaluation of models based on different PN techniques, and the simulation of the executable specification, which are used all together in a synergic manner.

We have illustrated how the separation between the graph-based structure of the model and dynamic properties such as the marking, enables the use of many structure-based analysis techniques. In order to verify some quantitative properties, namely throughput and economic cost, we have added a timing interpretation and have assumed an associated CPU pay-per-use cost. We have estimated performance and cost based on stochastic Petri nets, and conducted simulations under the light of our previous analysis covering the solutions space between the identified bounds. Besides, we made use of the wavefront algorithm, a simple Matrix-Vector multiplication in streaming fashion, as a recurrent example throughout the paper and in order to evaluate our proposal.

{\textcolor{black}{To sum up, this paper covers the gap between the capture of functional and non-functional requirements and the design specification given to developers using a specific platform or an architectural solution. However, our approach also comprises some limitations.
We believe that it is difficult to change current practice and its inertia, as our approach involves the incorporation of resources early at the functional specification. Additionally,
software engineers may not be familiar with formal methods. For these reasons, in order to foster its usage, we also provide a preliminary specification language to support our methodology, which includes high-level abstractions, and which due to space limitations can be found in the appendix.}

{\textcolor{black}{As future work, we plan to exploit our approach in a real cloud infrastructure in the terms exposed in Section~\ref{sec:exploit}. Our models can be used in
two different ways. On one hand, we can exploit the on-demand provisioning of cloud infrastructures in order to configure and manage the computational cloud resources
so that one of the task to resource mapping solutions studied can be arranged. On the other hand, the models can be the basis of an autonomic resource management approach, thereby the elasticity actions on the infrastructure are driven by the models obtained by the methodology. Besides, we also plan to study a formal definition of the Petri net subclass supporting the constructive process of the presented methodology, and we also need to design tools for assisting the realisation of our methodology.}

\section*{References}
\bibliographystyle{elsarticle-num}

\newpage

\section*{Appendix: Specification Language for basic and advanced data flow applications}
\label{sec:Language} 

The  principles presented in section~\ref{sec:mod} have guided the design of our component based  specification language to support the methodology for building trustworthy data intensive applications. We call this language specification  $Langliers$  (LANGuage of Layers and tIERS),  highlighting our approach to consider at least functional and operational level.  -The Hierarchical construction of the specification of CDFAs requires the definition of  composition and refinement primitives.  According to the methodology,  we start representing   the most basic building blocks   and continue with  their composition by means of simple operators. Composition operators provide the way to configure components with complex behaviours by means of {\em composites} and by the definition of a  {\em CDFA network} as a graph  showing various connected processing components that operates over a data stream. This processing network model is an abstraction that describes the {\bf functional} behaviour of a CDFA   made up of a number of platform-independent components.  The explicit network specification allows developers to visualize the functional model and apply analysis techniques.  
The  remaining steps to specify a complete model is to set the implementation of the functional model with the specification of the resources that will be used to execute the model.  %This {\bf operational} specification can be used to conduct performance  optimizations selecting a good mapping of Computational Processes and Data Transmission Processes to computational and network resources. 

\subsection*{Basic building blocks and composition operators}
\label{sec:primitives}

The precise and formal  specification of components requires a description beyond architectural units.  It is not enough to specify the interface. Specification of  nonfunctional or quality attributes  is also required to enable different types of  reasoning.  In $Langliers$ a component is  expressed in the form of an {\em interface} and a {\em behaviour} description. %\cite{Seceleanu:2013ol}.
The {\em \bf interface}  specifies the services the component provides,  and publishes  its input and output {\bf ports}.   It gives information at the syntax level that enables {\em data type checking}, and publishes the  {\em events} that trigger computations and state changes. This way, our data stream model follows a data driven execution model where events  lead to computations that may generate events on other components.  The {\bf \em  behaviour}  represents components internal states and state changes.   A component behavior description  is specified either by one explicit PN or by the  PN resulting from the behavior composition of  subcomponents.  It is inspired in digital systems VHDL language specification whose components specifications consist of a port declaration in the interface enumerating the events that change the component state, and the {\em architecture} declaration that describes either the entity's behavior or its structure.% \cite{Ashenden:2010ei}.

Once defined the elements of a component description, we present the basic building primitive components to describe a distributed system.  The  behaviour specification of these building primitives is represented by one of the PNs shown in Figure~\ref{Figure1}.  

Transitions and places used for the composition of behaviours are part of the interface declaration and can have respectively associated a set of attributes. For example, places have predefined attributes such as type and initial marking, transitions have predefined attributes such as time or  actions, and arcs have label attributes.  Hence,  behaviour specification can have  inscription expressions labelling arcs in the same way as high-level PNs (HLPNs) %\cite{Jensen:1991aa} 
(e.g., colored PNs,  predicate/transition nets, or Object Petri Nets), which provides more compact and manageable descriptions.  Internal transitions with  labelled  actions represent atomic actions that may be executed after the firing of the transition.  In this way, any conventional programming language can be used to describe operations and provides an executable functional specification. It is similar to the  embedded user defined functions of data intensive engines (e.g Spark, Storm, Flink), subprograms in VHDL specifications, or the use of Java expressions in    {\tt Renew}.
Textual representation of components  follows a frame-based tradition of concepts as attribute-value pairs. The component specification is started with  the component's name, and follows with a {\tt Interface}  list of place and transitions that can be used to compose components. The  {\tt Behaviour} specification is equivalent to the {\tt architecture} specification of a VHDL component. The behaviour specificaiton begins with a list of {\tt components}: either references to  $PN$s  subcomponents, {\tt Links} that connect components,  and {\tt Rename} declarations;  {\tt Places} and {\tt Transitions}  use a dotted notation to  reference the component name followed by the Place/Transition name; and places and transitions attributes can be referred by the name of place/transition followed by a dot and the attribute's name (see Figures~\ref{Figure7} and \ref{Figure8}) . 

%For the sake of simplicity, we will focus on the specification of components and the  process of stepwise refinement from one level of abstraction to the next lower level. For this reasons, PN specification are presented graphically, and the textual specifications of components will point by means of labels to PNs or to the PN elements that are part of the interface or behavior.    

%Note that the {\bf Links} declaration can be complex enough to require the use of a procedural language for the composition of operators.  We must not forget that the main purpose is the specification of parallel abstractions. I Therefore, a declarative specification of the regular structure will facilitate the specification of a wide range of problem sizes. %Mover al modelo functional

%~\cite{DBLP:conf/apn/KummerWDSKMRV04}
\begin{figure}
	\includegraphics[width=0.8\textwidth]{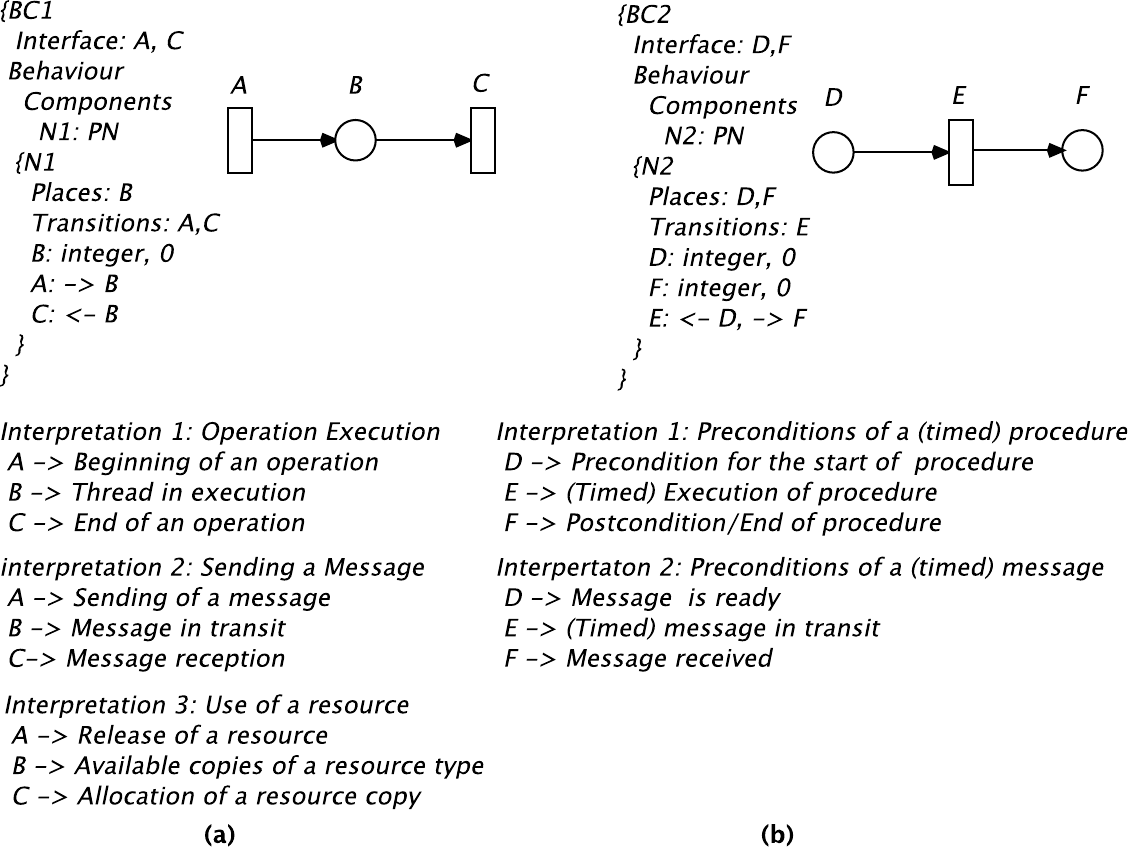}
	\centering
	\caption{Component specification of basic component primitives and interpretations.}
	\label{Figure8}
\end{figure}

We will start with the control specification, which can be subsequently enriched with data, combining in this way the control and data-flow perspectives of the model.
For control model, we propose two basic  building-blocks with different precise interpretations. Figure~\ref{Figure8}(a) shows the  basic component $BC_{1}$ that declares as behaviour the Petri net $\N_{1}$.  Interface Transition  $A$ requires a component with a Transition (output port) that triggers transition $A$. Transition $A$  represents the beginning of the execution of an operation in the first interpretation, the beginning of the sending of a message in the second interpretation and the release of a resource in the third interpretation.  Interface Transition $C$  requires a component to trigger the execution of an event synchronised with the end of the operation in the first interpretation, the reception of a message in the second interpretation and the allocation of a resource copy in the third interpretation.  Place $B$ in Figure \ref{Figure8}(a) is an internal state with the  interpretations presented.   The type attribute specifies the type $token$ which means black tokens that does not carry data information, and the. The  initial marking by defaults specify zero tokens, but this attribute can be specified when the  component is declared as subcomponent of another component.  Arc transitions are specified.   Attribute labels are empty denoting that one black token is respectively removed/added from previous/posterior place.

Figure~\ref{Figure8}(b) shows the $BC_{2}$ that declares as behaviour the PN $\N_{2}$. Interface Place  $D$  represents events received and requires a component that sends events to this place, and Interface Place $F$  represents generated events waiting to be consumed requiring a component that consumes these events.  Place $D$ represents preconditions to start the execution of an operation in the first interpretation and that a message is ready in the second interpretation. Place $F$  represents the ending of an operation or a received message waiting to be consumed.  Transition $E$ represents an internal action,  for example a procedure executed in a machine or a network transmission. It can be timed to represent non-instantaneous actions. 

\subsection*{Composition operators}

The declaration of a "{\em Composite}" component is similar to a basic component   with a  declaration in the behavior that consists of subcomponents and/or $PN$s, connections between them, and the definition of  its interface  as a mapping of input and output ports to the input and output ports of subcomponents. Three simple composition operators are provided to define Composites. They are  based on the fusion, split and copy of transitions and places in the interface: {\bf \tt split}, {\bf \tt fusion} and {\bf \tt copy}. The {\bf \tt split} operator replaces the argument, place or transition, by  a couple of places or transitions, where the first resulting place or transition has the input arcs of the split argument, and the second one the output arcs.  The {\bf \tt fusion} operator replaces the arguments, places or transitions, by a new place or transition that has the input and output arcs of the arguments. Finally, the {\bf  \tt copy} operator creates a new place or transition identical to the argument. 

The construction of the functional model is based on the identification of the basic components that configures a CDFA network:  CPs and DTPs. Figure~\ref{Figure9} shows the $Langliers$ specification of an untimed and timed  CPs. Figure~\ref{Figure9}.(a) shows the component $Op$ that specifies  a CP. This basic component is defined by the composition of two instances of $BC_1$, the first one with the first interpretation and the second with the third interpretation semantic. The second component declaration includes in the constructor declaration an initial marking of one token in Place $B$.  On the right is represented the composition of PNs describing the subcomponents behaviour, and on the left the textual component specification.   Composites consist in a declaration of subcomponents, and a declaration of port mappings  or links by means of the PN based composition operators. The rename  declaration is used to rename places and transitions of subcomponents or the  result of composition operators.

\begin{figure}
	\includegraphics[width=1\textwidth]{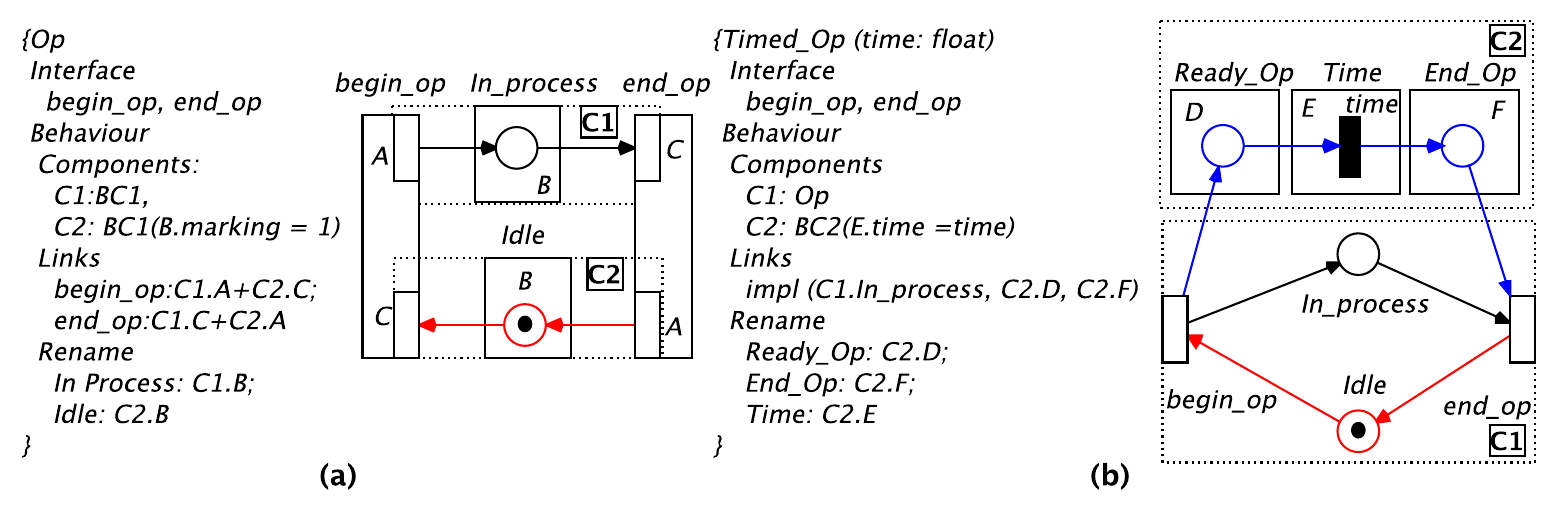}
	\centering
	\caption{A Computational Process with Resources. a) A CP composed by  1 state. b) Timed model assigning $time_i$ units of time to the execution of the $operation_i$. }
	\label{Figure9}
\end{figure}

A step-wise refinement of this component considering operational aspects can be developed by the  addition of timed information to a CP. It is introduced by the addition of a sequence place-transition-place in parallel with a process place representing an operation of the computational task that consumes time. The transition added is labelled with time information representing the duration of the computational operation. In Figure~\ref{Figure9}.(b), the CP from Figure~\ref{Figure9}.(a) is refined by assigning \emph{time}  units of time to the execution of the operation. The constructor receives as argument the attribute time value of  $E$.  For operational refinements two new operators are defined  using basic operators.   {\bf \em Split-Copy} splits a copy of a place, and {\bf \em impl} operator specifies the way an operation is implemented. The  {\bf \em Split-Copy}  operation can receive as argument an attribute label for each new arc.  The  {\bf \em impl} can have three arguments. In the first case, it makes a {\bf \em split-Copy} of the first argument and fusions the first returned place with the second argument, and the second place with the third argument.  In the second case, it makes a fusion of first with third, and second with fourth argument. The {\tt Links} declaration in Figure~\ref{Figure9}.(b) uses the   {\bf \em impl} operator to refine $In\_process$.

\subsection*{Specification enrichment with data and object oriented features}

\begin{figure}
	\includegraphics[width=1\textwidth]{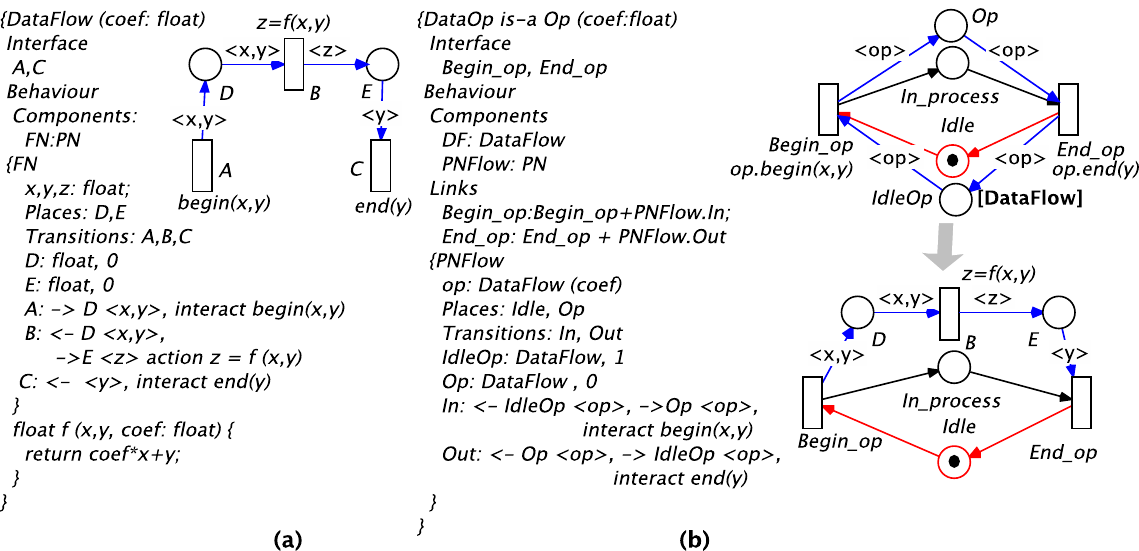}
	\centering
	\caption{Specification enrichement with data and object .}
	\label{Figure10}
\end{figure}

Once the modeler specifies the  control level of the the functional model, it is possible the enrichment by means of the incorporation of data, functions, and guards using HLPNs in the model. It allows the separation of control and data-flow specifications. 

Additionally to data, we incorporate object oriented features to the specification language with a double purpose. On the one hand, we introduce the idea of simple inheritance  in order to code reuse. In this way we can reuse one specification adding new subcomponents/PNs to the inhered behaviour (architecture).  On the other hand, tokens in a Petri net can be interpreted as PNs. In this way we can model software modules  as agents that does not belong to a specific environment but are also able to switch to a different one.  This is the main motivation of the definition of Object Petri nets using the Nets-within-Nets paradigm. % \cite{Valk:2004qf}. 
In the Nets-within-Nets paradigm a token net represents a task, with a   plan for the execution procedure, that can be executed in different machines or locations. Token nets are called {\em object net}s in distinction to the {\em system net} to which it belongs.  Interactions between objects nets, or between the object nets and the system nets are represented by labels as will be illustrated in the samples.

%Comentar semantica  y que el modelo permite razonar sobre la correccin semantica. 
Figure~\ref{Figure10}.(a) represents a simple plan modelled by a PN.  The DataFlow declaration introduces the requirement of providing an argument in the constructor. It specifies the label attributes of all arcs,  the function $f$ that label Transition $B$, and   the attribute {\tt interact} in transitions $A$ and $C$ that specifes the interaction of this {\em object net} with a {\em system net}.  The declaration of function $f$ is done in a C/Java style, but any other language can be used. Figure~\ref{Figure10}.(b) specifies the $DataOp$ component by inheritance of all components of a $Op$  component  in Figure~\ref{Figure9} that is refined by the composition of the inherited net with the specified $PNFlow$, which complete it with the data-flow.  $PNFlow$ has one instance of $DataFlow$ as initial marking of the Place $IdleOP$.  In this case, $PNFlow$ is the system-net, and $DataFlow$ the object-net.  

The bottom PN of Figure~\ref{Figure10}.(b) shows the equivalent PN. It is important to note that we have borrowed the Renew syntax for Object nets.  However, it is important to note that in order to avoid the semantics problems that can introduce distributed tokens semantics, that is,  tokens of those object nets whose copies are distributed to different output-places we will  avoid more than one copy in different places. In this way we avoid the ubiquity problem until a Petri net formalism can support it with a clear semantic and providing valid analytical results.   
%Figura y describir el ejemplo.

%\subsection{$Langliers$ specification  of a Cloud based Operational Model of a Matrix-Vector Multiplication in streaming}
\subsection{$Langliers$ specification  of untimed functional  specification of the wavefront algorithm} 
\label{sec:wavefront}
\begin{figure}[h]
	\includegraphics[width=0.9\textwidth]{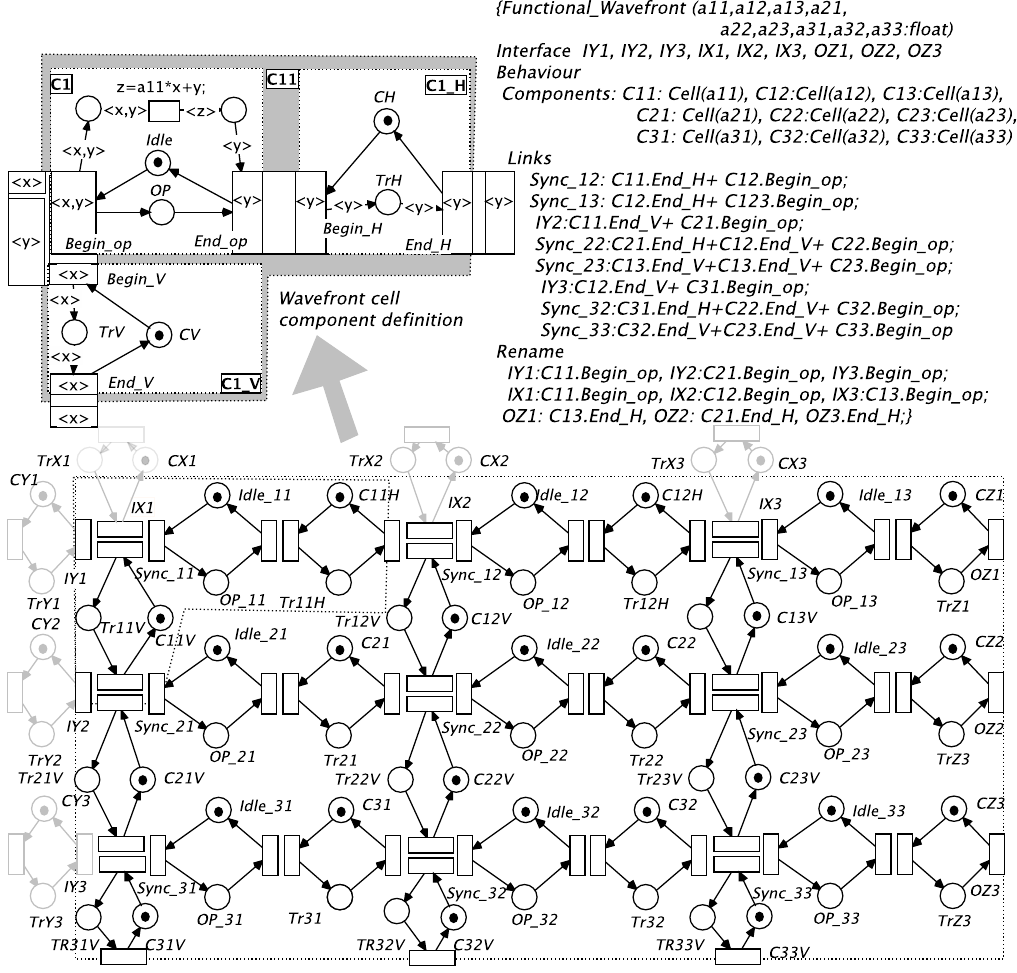}
	\centering
	\caption{Functional $3 \times 3$ wavefront array.}
	\label{Figure11}
\end{figure}

The functional  model is constructed in a modular fashion.  Each element of the wavefront  array is defined as a composite $Cell$ component  made up of basic subcomponents (see upper  Figure~\ref{Figure11} $Cell$ component specification): (1) $C1$,  a subcomponent to describe the CP carried out in a node of the wavefront array; and  (2) $C1\_H$ and $C1\_V$, two subcomponents to describe DTPs of the  data streams from $Cell$s at the  north/east to $Cell$s in the south/west.   Observe the $C1$ component is the component specified in  Figure~\ref{Figure10}.  Transition   labeled with the function $z=aij*xj+yi$  specifies the action that must be carried by each $Cell$ component. Each $Cell$ component is initialised with its coefficient $aij$ element of matrix $A$. Arcs and transitions are annotated with variables defining a complete functional specification. 

To construct the global functional model, nine instances of the $Cell$ component are needed (see Figure ~\ref{Figure11}.).  Each $Cell_{ij}$ component  is connected by a $Sync\_ij$ transition  resulting from the fusion of its transition $begin\_op$ with  transition  $End\_H$ of the east  $Cell_{i(j-1)}$; and its transition $End\_V$ with transition $begin\_op$ of the south $Cell_{(i+1)j}$ .   Each one of $Sync\_1j$   transition at the first row does not have a $Cell$ component connected at the North, and  define the $IYi$ interface transitions  representing the input stream of the corresponding i-th component of the vector $X^{(k)}$ % via the fusion of the transitions $begin\_op\_1j$ with a the $end\_Transmission$ of a DTP component. 
In the same way, each one of  $Sync\_i1$  transition   of the first column does not have a left $Cell$ component, and constitute the  $IXj$ interface transitions  representing the input stream of the corresponding j-th component of the vector $Y^{(k)}$. % via the fusion of the transitions $begin\_op\_i1$  with a the $end\_Transmission$ of a DTP component. 
Each  $End\_H$ transitions of $Cell_{i3}$ components in the last column is the output stream of the corresponding i-th component of the vector $Z^{(k)}$, and defines the $OZi$ interface transition. This models constitute a complete operative functional model, which can be interpreted/executed injecting data and obtaining the results.

\subsection*{$Langliers$ specification  of a Cloud based Operational Model of the wavefront algorithm} 
\label{sec:wavefrontOp}

\begin{figure}[h]
	%\vspace{7cm} \hspace{1.5cm} \special{picture PNWF scaled 1000}
	\includegraphics[width=1\textwidth]{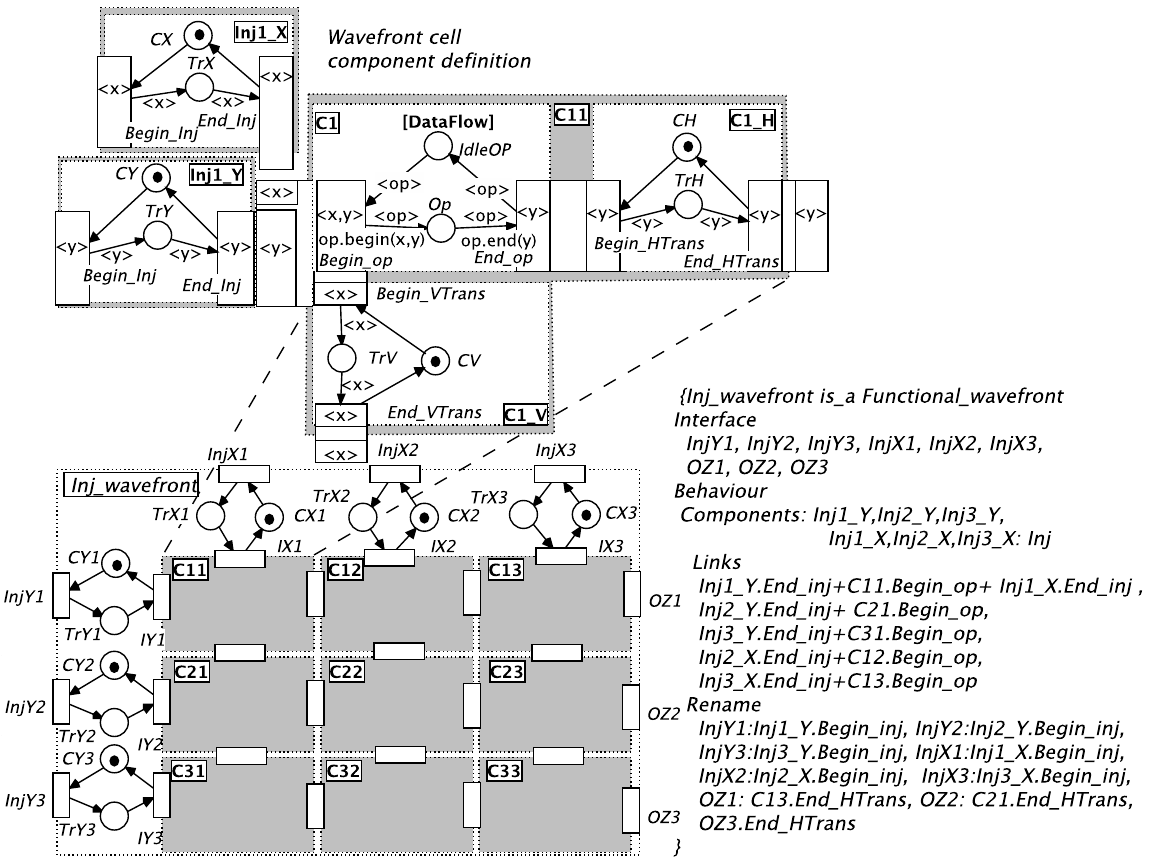}
	\caption{Operational  $3 \times 3$ wavefront array executed with injectors.}
	\label{Figure12}
\end{figure}

We assume a cloud platform (OpenNebula, OpenStack, etc.) that  provides networking and computing virtualisation. For each token in the Idle places $Idle\_ij$ we assume that there is a VM implementing a computational resource that can accomplish the same functionality with similar performance, and for each token in the $CijH$ and $CijV$ places we assume a transmission over the network assuming than the use of virtualisation technologies does not alter the original incoming data injection rate at each data flow.  To illustrate the step-wise  refinement of the functional,   we start adding injectors components to the $Functional\_Wafefront$ presented in Figure~\ref{Figure11} .  The aim of these components is to regulate the injection process to avoid the variance  of injection rate.  We can model this mechanism, for example, by means of a token-based injection mechanism. Figure~\ref{Figure12} shows the $I nj_wavefront$ component, which incorporate the injection regulation components. Observe that in this figure we have used the object net version of the data/functional specification, with $DataFlow$ as marking of places $IdleOP$.

\begin{figure} [h]     
	%\vspace{7cm} \hspace{1.5cm} \special{picture PNWF scaled 1000}
	\includegraphics[width=1\textwidth]{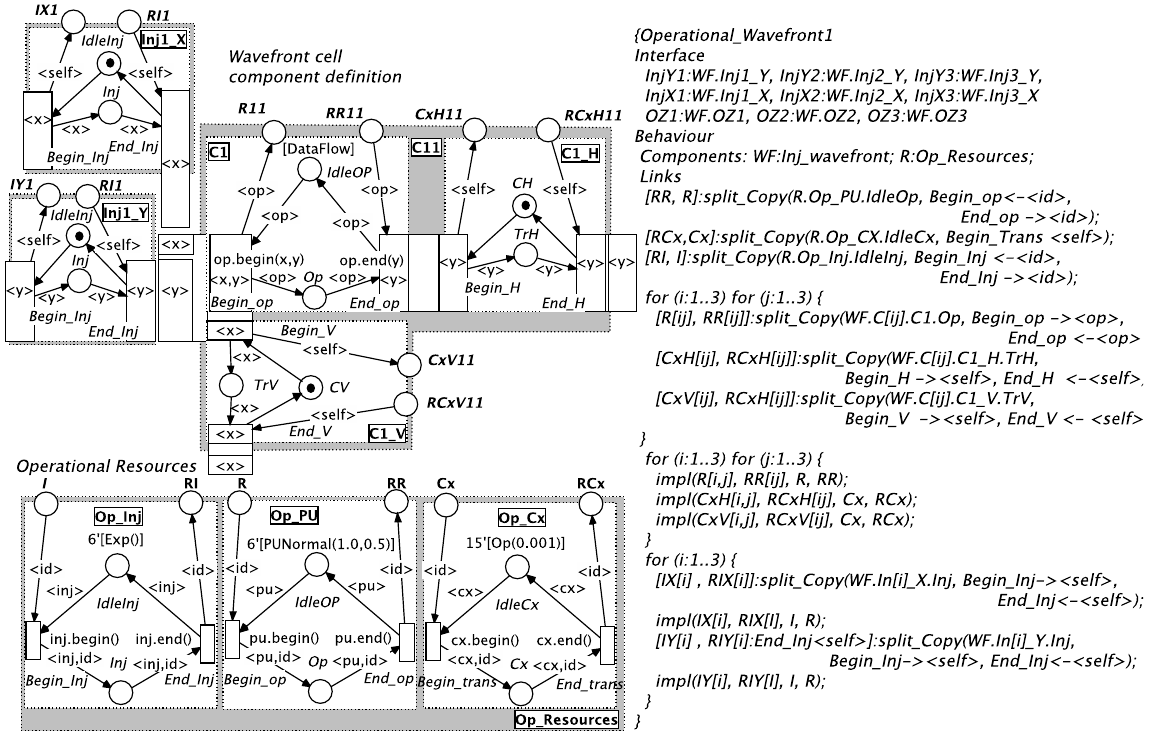}
	\centering
	\caption{Operational wavefront array executed over nine independent VMs.}
	\label{Figure13}
\end{figure}

The addition of time to  the model of Figure~\ref{Figure12} can be done in the way described in the previous sections: adding a sequence place-transition-place in parallel with the places representing the activities that consume  time  by the $Cell$ located at row i-th, column j-th: $Op$ representing the duration of the computation, and   in parallel with places $TrH$ and $TrV$   representing the consumption of time in the transmission of a data element in the corresponding DTPs.  An alternative way is to refine the functional model with a specification of the resources and introducing the time in a model of resources closer to the target platform. Figure~\ref{Figure13} shows this approach. The $Operational\_Wavefron1$ has two components, the functional model specified in  Figure~\ref{Figure12} , and the specification of resources for computing, transfer data, and injection.  The {\em mapping} of resources is specified by means of place fusion.  The {\tt Link} part of the specification  generates the places $I$, $RI$, $R$, $RR$, $Cx$ and $RCx$,  which are respectively  the  result of the, {\tt split-copy} of places $Inj$, $Op$ and $Cx$ of the $Op\_Resources$ subcomponent. Places $R_{ij}$, $RR_{ij}$,  $CxH_{ij}$, $RCxH_{ij}$, $CxV_{ij}$, $RCxV_{ij}$ are respectively  the {\tt split-copy} results of places $Op$, $TrV$ and $TrH$; and $IX_{i}$, $RI_{i}$, $IY_{i}$, $RI_{i}$ are respectively  the {\tt split-copy} results of places $Inj$ of the $Inj\_wavefront$ subcomponent. Each $R_{ij}$, $RR_{ij}$ is respectively fused with places $R$ and $RR$. Note that all components share the same $R$ and $RR$ place to obtain/release a resource to execute the operation. To distinguish the  beginning and ending of each operation,  the reference to the $DataFlow$ instance is passed to the $Op\_PU$ subcomponent.  In this way, each operation/Resource binding is uniquely identify.
The same process is followed by  DTPs and injectors.  In this case, arcs labeling  arcs from/to the result of split places have the $<self>$ reference, which is a reference to the subcomponent instance with the same purpose of identifying the functional component that is interacting with resources.

The bottom of Figure~\ref{Figure12} shows  the resource specification, which have only  information of different probability distributions with simulation purposes. It is  assumed that  resources are provided without any dependency between them; injection follows an exponential distribution;  service time follows a normal distribution (mean 1.0 second,  typical deviation 0.5); and  transfer of data is negligible (1 mlsec).  Following this approach we obtain an operational  net model that is isomorphous to the functional model, and we can perform the quantitative analysis presented in section~\ref{sec:analysis}. 
\begin{figure}[h]
	\includegraphics[width=1\textwidth]{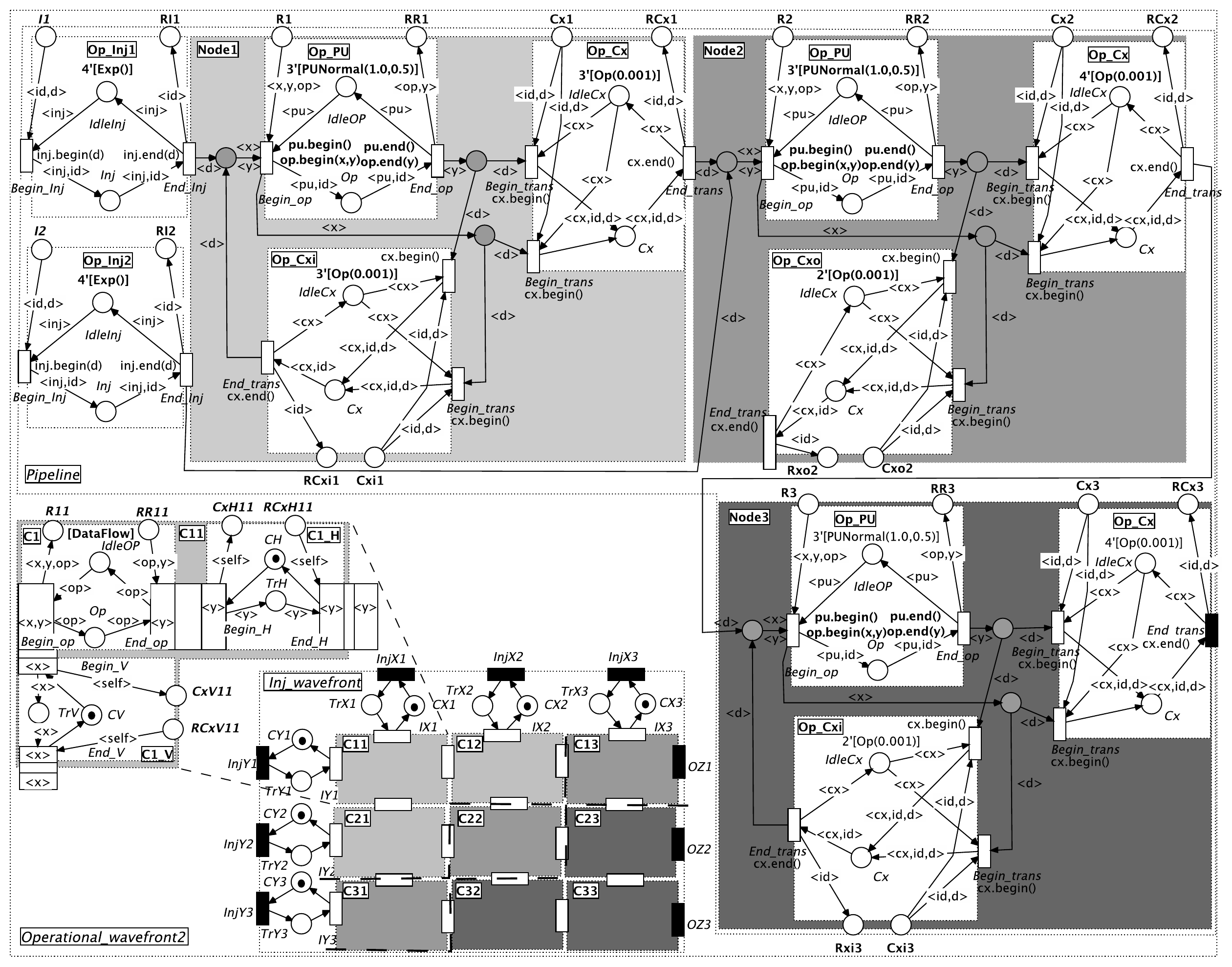}
	\centering
	\caption{Operational streaming $3 \times 3$ wavefront array executed in a Pipeline.}
	\label{Figure14}
\end{figure}

A step forward an operation model  closer to the implementation is shown in Figure~\ref{Figure14}. It shows a refined operational model considering the {\em algorithm structure design space}. In this case, we have chosen  a pipeline of three nodes to process the wavefront in streaming. Observe that the separation of functional and operational specifications allows modeler to  reuse  the functional specification for a new operational model.  Each node component is composed of: (1)   $OP\_PU$, a subcomponent to describe the available Computational Resources  at each node, and  (2)  two subcomponents to describe DTPs between computational resources in the same node, and  to the computational resources of the next step.   Observe that we have refined the cell components using the object oriented version. The cell component has an initial object token marking $[DataFlow]$.   Another important difference of  the  operational component $Operational\_wavefront2$ is that  it  translates the interaction with the $DataFlow$ object net to the $Op\_PU$ components of the pipeline nodes. In this ways, the pipeline represents the data-flow and the movement of the code to the processing resources. It is represented by the {\tt op.begin(x,y)} and {\tt op.end(y)} interactions in  $OP\_PU$.

Each $Node_{i}$ component has an $OP\_PU$ component with an $Op$ place representing the VM resources in use, and an $IdleOp$ place representing available VMs allocated for this step.  The marking $3'[PUNormal(1.0,0.5)]$ of places $OP\_PU$ specifies an initial marking of three objects nets that are used to simulate computational resources whose service time  follow a normal distribution (mean 1 second, typical deviation 0.5). In the same way $Op\_Cx$ components represents transfer of data between resources of different nodes  or with external components, and $Op\_Cxi$ transfer of data  between resources in the same node.     Observe that the  intermediary node  $Node_{2}$ does not contain transfer of data between internal resources, and that $Op\_Cxo$ represents data transfer  to external  components. 

$Operational\_Wafefront2$  refines the $Functional\_wavefront$  component in the same way splitting the same places of the functional model.  The only difference is the mapping with the resources of the nodes in the pipeline. CPs in a left to right diagonal can operate concurrently and propagate in wavefronts. For this reason, a balanced deployment of the wavefront $Cell$ components over the pipeline steps is deployed in the following way: In the first node, $Cell11$,  $Cell12$ and $Cell21$ that corresponds from left to right to the   first and second diagonal.  $Cell$s in the middle diagonal are executed in the second step, and the rest of $Cell$s are executed in the last node. The horizontal connection of the $Cell11$ is an internal transmission between VMs executing in the same node, while its vertical connection is a transmission with a VM in the second node.  In the case of the $Cell22$, all its connections are external transmissions, and the same happens with the $Cell$s in the diagonal deployed in the middle step.  Data injection is done in cells mapped to the the first and second node.  The textual component specification of this mapping is presented in  Figure~\ref{Figure15}.

\begin{figure}[h]
	\includegraphics[width=1\textwidth]{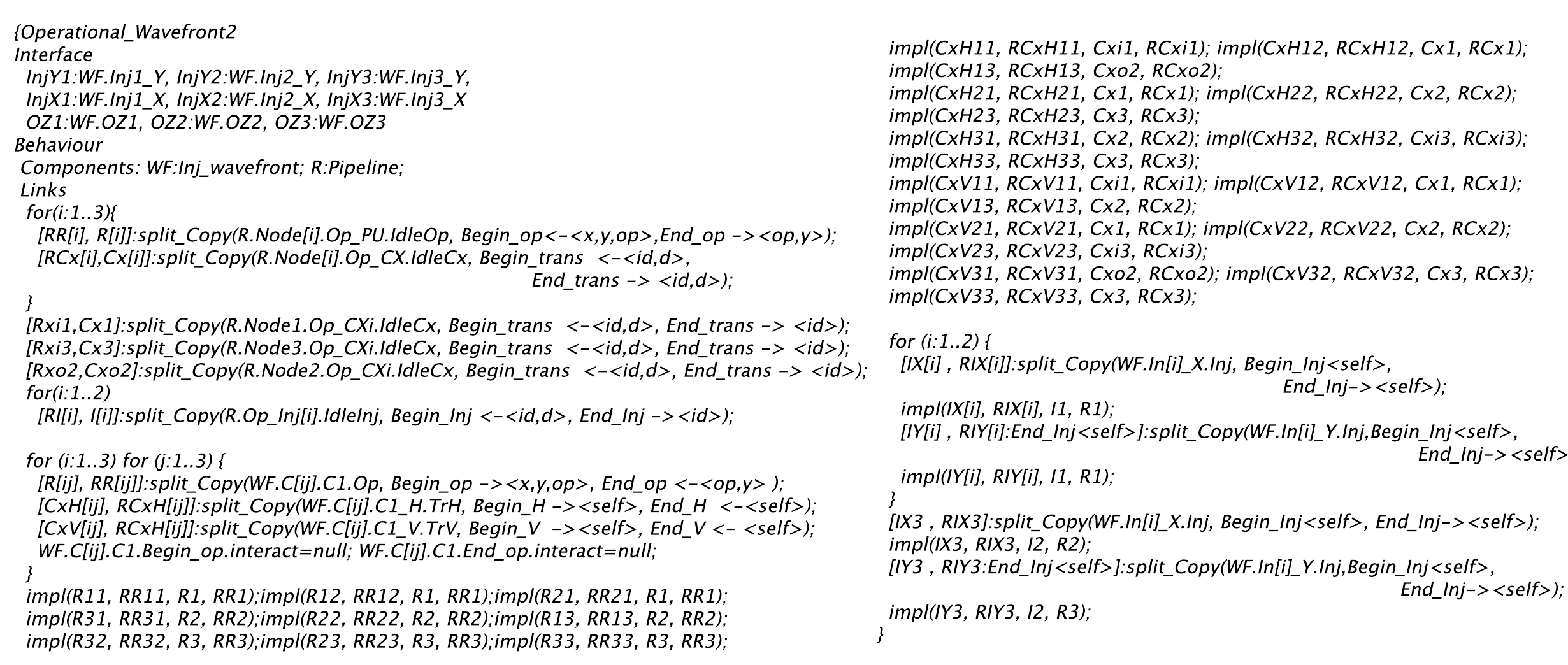}
	\centering
	\caption{Textual $Langliers$ specification in  Figure~\ref{Figure14}.}
	\label{Figure15}
\end{figure}

\end{document}